\documentclass[a4paper,12pt, epsfig]{article}
\def\letter{0}\def\pr{0}
\usepackage{epsfig}
\usepackage{epstopdf}
\usepackage{textcomp}
\usepackage{graphicx}
\usepackage{ifthen}
\usepackage{ulem}

\usepackage{amsmath}
\usepackage[psamsfonts]{amssymb}
\usepackage{euscript}

\usepackage{latexsym}
\usepackage[arrow,matrix,curve]{xy}

\usepackage[hypertexnames=false]{hyperref}
\hypersetup{
    colorlinks=true,
    linkcolor=blue,
    filecolor=magenta,   
    citecolor=black,   
    urlcolor=cyan,
    }

\newskip\humongous \humongous=0pt plus 1000pt minus 1000pt

\newif\ifdtup

\def\,{\hspace{-.1cm}}
\def\hsp{,\hspace{.7cm}}

\def\dvx{\int d^3\vx}

\def\fc#1#2 {\frac{n}{q}#1\frac{n}{q}#2}

\newcommand{\vac}{\ensuremath{|0\rangle}}
\newcommand{\ovac}{\ensuremath{|\Omega\rangle}}

\renewcommand{\tanh}{\textrm{tanh}}
\newcommand{\sech}{\textrm{sech}}
\newcommand{\csch}{\textrm{csch}}

\def\exp#1{\hbox{\rm exp}\left[#1\right]}

\renewcommand{\theequation}{\arabic{section}.\arabic{equation}}
\renewcommand{\(}{\begin{equation}}
\renewcommand{\)}{end{equation} \vspace{-.05in}\linebreak}

\newcounter{saveeqn}
\newcounter{savealpheqn}

\newcommand{\alpheqn}{\setcounter{saveeqn}{\value{equation}}%
  \stepcounter{saveeqn}\setcounter{equation}{0}%
  \renewcommand{\theequation}{\mbox{\arabic{section}.\arabic{saveeqn}
\alph{equation}}}
  \renewcommand{\)}{\end{equation}}}
\def\part#1{\frac{\partial}{\partial{#1}}}%
\def\group#1{\refstepcounter{equation}\setcounter{saveeqn}
 {\value{equation}}%
  \label{#1}\setcounter{equation}{0}%
\renewcommand{\theequation}{\mbox{\arabic{section}.\arabic{saveeqn}
\alph{equation}}}
  \renewcommand{\)}{\end{equation}}}
\newcommand{\reseteqn}{\setcounter{equation}{\value{saveeqn}}%
  \renewcommand{\theequation}{\arabic{section}.\arabic{equation}}%
  \renewcommand{\)}{\end{equation}}}

\newcommand{\aalpheqn}{\setcounter{saveeqn}{\value{equation}}%
  \stepcounter{saveeqn}\setcounter{equation}{0}%
  \renewcommand{\theequation}{\mbox{
        \Alph{subsection}.\arabic{saveeqn}\alph{equation}}}
   \renewcommand{\)}{\end{equation}}}
\newcommand{\areseteqn}{\setcounter{equation}{\value{saveeqn}}%
  \renewcommand{\theequation}{\Alph{subsection}.\arabic{equation}}%
  \renewcommand{\)}{\end{equation}}}

\renewcommand{\thefootnote}{\alph{footnote}}
\renewcommand{\(}{\begin{equation}}
\renewcommand{\)}{\end{equation}}
\newcommand{\ba}{\begin{eqnarray}}
\newcommand{\ea}{\end{eqnarray}}

\renewcommand{\sl}{{\sqrt{\lambda}}}

\newcommand{\cbp}{\mathop{\vtop{\ialign{##\crcr
   $\hfil\displaystyle{}\hfil$\crcr\noalign{\kern-13pt\nointerlineskip}
   \BIG{)}\hskip0pt\crcr\noalign{\kern3pt}}}}}
\newcommand{\pa}{\mathop{\vtop{\ialign{##\crcr

$\hfil\displaystyle{\oplus}\hfil$\crcr\noalign{\kern+1pt\nointerlineskip
}
   \hspace{.08in}$^{\alpha=0}$\hskip6pt\crcr\noalign{\kern3pt}}}}}
\renewcommand{\hsp}{,\hspace{.3in}}
\newcommand{\p}{^\prime}

\catcode`\@=11
\def\vereq#1#2{\lower3pt\vbox{\baselineskip1.5pt \lineskip1.5pt
\ialign{$\m@th#1\hfill##\hfil$\crcr#2\crcr\sim\crcr}}}
\catcode`\@=12

\renewcommand{\(}{\begin{equation}}
\renewcommand{\)}{\end{equation}}

\def\vx{{\vec{x}}}
\def\vp{{\vec{p}}}
\def\vpp{{\vec{p}^{
\hspace{.05cm}\prime
}}}
\def\vk{{\vec{k}}}
\def\vkp{{\vec{k}\p}}

\def\pin#1{\int \frac{d#1}{2\pi}}
\def\ppin#1{\int\hspace{-17pt}\sum \frac{d#1}{2\pi}}
\def\ppink#1{\int\hspace{-17pt}\sum\frac{d^{#1}\vk}{(2\pi)^{#1}}}

\def\ppinkp#1{\int\hspace{-17pt}\sum\frac{d^{#1}\vk\p}{(2\pi)^{#1}}}
\def\dint{\int\hspace{-12pt}\sum }

\def\pinv#1#2{\int \frac{d^{#1}\vec{#2}}{(2\pi)^{#1}}}
\def\ppinv#1#2{\int \frac{d^{#1}\vec{#2}^{
\hspace{.07cm}\prime
}}{(2\pi)^{#1}}}

\def\Ad#1{A^\ddag_{\vp_{#1}}}

\def\df{\mathcal{D}_{f}}
\def\dv{\mathcal{D}_{v}}

\usepackage[dvipsnames]{xcolor}

\newcommand{\beas}{\begin{eqnarray*}}
\newcommand{\eeas}{\end{eqnarray*}}

\newcommand{\bquo}{\begin{quote}}
\newcommand{\enqu}{\end{quote}}

\def\lim#1{\stackrel{\rm{lim}}{{}_{#1}}}

    \newcommand{\g}{\mathfrak g}

\def\ch{{\mathcal{H}}}

\def\ok#1{\omega_{k_{#1}}}

\def\ovp#1{\omega_{\vp_{#1}}}
\def\ovpp#1{\omega_{\vpp_{#1}}}

\def\V#1{V^{(#1)}(\sqrt{\lambda}f(x))}

\newcommand{\beq}{\begin{equation}}
\newcommand{\eeq}{\end{equation}}
\newcommand{\bea}{\begin{eqnarray}}
\newcommand{\eea}{\end{eqnarray}}

\newskip\humongous \humongous=0pt plus 1000pt minus 1000pt

\newif\ifdtup

\catcode`\@=11

\ifthenelse{\equal{\letter}{0}}{ %se lettere=0, commenta questo

\@addtoreset{equation}{section}
\def\theequation{\arabic{section}.\arabic{equation}}

\def\@normalsize{\@setsize\normalsize{15pt}\xiipt\@xiipt
\abovedisplayskip 14pt plus3pt minus3pt%
\belowdisplayskip \abovedisplayskip
\abovedisplayshortskip \z@ plus3pt%
\belowdisplayshortskip 7pt plus3.5pt minus0pt}

\def\small{\@setsize\small{13.6pt}\xipt\@xipt
\abovedisplayskip 13pt plus3pt minus3pt%
\belowdisplayskip \abovedisplayskip
\abovedisplayshortskip \z@ plus3pt%
\belowdisplayshortskip 7pt plus3.5pt minus0pt
\def\@listi{\parsep 4.5pt plus 2pt minus 1pt
      \itemsep \parsep
      \topsep 9pt plus 3pt minus 3pt}}

\relax

\def\section{\@startsection{section}{1}{\z@}{3.5ex plus 1ex minus  .2ex}{2.3ex plus .2ex}{\large\bf}}

\def\thesection{\arabic{section}}
\def\thesubsection{\arabic{section}.\arabic{subsection}}

\def\appendix{\setcounter{section}{0}
 \def\thesection{Appendix \Alph{section}}
 \def\thesubsection{\Alph{section}.\arabic{subsection}}
 \def\theequation{\Alph{section}.\arabic{equation}}}
\renewcommand{\theequation}{\arabic{section}.\arabic{equation}}

}{
\renewcommand{\theequation}{\arabic{equation}}

} %se letter=0, commenta questo

\begin{document}
% ========================================================================
\def\thefootnote{\fnsymbol{footnote}}
\def\thetitle{Constructing A Finite Tension Domain Wall in  $\phi^4_4$}
\def\autone{Jarah Evslin}
\def\autthree{Hui Liu}
\def\autfour{Baiyang Zhang}
\def\auttwo{Hengyuan Guo}
\def\affd{Yerevan Physics Institute, 2 Alikhanyan Brothers St., Yerevan, 0036, Armenia}
\def\affb{University of the Chinese Academy of Sciences, YuQuanLu 19A, Beijing 100049, China}
\def\affa{Institute of Modern Physics, NanChangLu 509, Lanzhou 730000, China}
\def\affc{School of Physics and Astronomy, Sun Yat-sen University, Zhuhai 519082, China}
\def\affe{Institute of Contemporary Mathematics, School of Mathematics and Statistics,Henan University, Kaifeng, Henan 475004, P. R. China}
%\def\affc{Departamento de F\'isica de Part\'iculas, Universidad deSantiago de Compostela and Instituto Galego de F\'isica de Altas Enerxias% (IGFAE), E-15782
%Santiago de Compostela, Spain}
%\def\affd{Physics Dept., Brookhaven National Laboratory, Bldg. 510A, Upton, NY 11973, USA}
%\def\affc{Wigner Research Centre, H-1525 Budapest 114, P.O.B. 49, Hungary}

%\title{Titolo}

\ifthenelse{\equal{\pr}{1}}{
\title{\thetitle}
\author{\autone}
\author{\auttwo}
\author{\autthree}
\affiliation {\affa}
\affiliation {\affb}
\affiliation {\affc}
%\affiliation {\affd}
%pr e uno

}{

\begin{center}
{\large {\bf \thetitle}}

\bigskip

\bigskip

%\catcode`@=11

{\large 
\noindent  \autone{${}^{12}$}
\footnote{jarah@impcas.ac.cn},
\auttwo{${}^{3}$}
\footnote{guohy57@mail.sysu.edu.cn }, %\footnote{kehindeoogundipe@gmail.com},
\autthree{${}^{4}$} \footnote{hui.liu@yerphi.am}, 
\autfour{${}^{5}$\footnote{byzhang@henu.edu.cn}}
}

%{\large \noindent  \autone{${}^{1,2}$} \footnote{jarah@impcas.ac.cn} and \auttwo{${}^{1,2}$} \footnote{guohengyuan@impcas.ac.cn}}

\vskip.7cm

1) \affa\\
2) \affb\\
3) \affc\\
%4) School of Physics and Astronomy, Sun Yat-sen University, Zhuhai 519082, China\\
4) \affd\\
5) \affe\\

\end{center}

}

\begin{abstract}
\noindent
%We consider the domain wall in the (2+1)-dimensional $\phi^4$ double-well model, created by extending the $\phi^4$ kink in an additional infinite direction.  Classically, the tension is $m^3/3\lambda$ where $\lambda$ is the coupling and $m$ is the meson mass.  At order $O(\lambda^0)$ all ultraviolet divergences can be removed by normal ordering, less trivial divergences arrive only at the next order.  This allows us to easily quantize the domain wall, working at order $O(\lambda^0)$.  We calculate the leading quantum correction to its tension as a two-dimensional integral over a function which is determined analytically.  This integral is performed numerically, resulting in $-0.0866m^2$.  This correction has previously been computed twice in the literature, and the results of these two computations disagreed.  Our result agrees with and so confirms that of Jaimunga, Semenoff and Zarembo.  We also find, at this order, the excitation spectrum and a general expression for the one-loop tensions of domain walls in other scalar models.

\noindent
We have recently claimed that the domain wall in the 3+1 dimensional $\phi^4$ double-well model can be constructed as a squeezed, coherent state and that at one loop it has a finite tension given general, but unspecified, renormalization conditions.  In the present note, we justify this claim by showing that the tadpole is finite and the infrared divergences cancel exactly.  Also we carefully treat the renormalization of the normal ordering mass scale.  Faddeev and Korepin have stressed that ultraviolet divergences cancel in the soliton sector if they cancel in the vacuum sector when the corresponding calculations are identical in the ultraviolet.  We therefore renormalize the divergences in the vacuum sector using a Schrodinger picture prescription, which mirrors closely the analogous calculations in the domain wall sector.
\end{abstract}

%In 1+1 dimensions, it is well known that the quantum states corresponding to solitons are well described by coherent states.  In his 1975 Erice lectures, Coleman observed that this construction does not extend to higher dimensions, as the coherent states have infinite energy density.  He challenged the students to construct the quantum states corresponding to solitons in higher dimensions, a problem which remains unsolved today.  However, %As emphasized recently by Berezhiani, Cintia and Zantedeschi,
%even in 1+1 dimensions the correct quantum states are actually given by deformations of coherent states.  In the 3+1 dimensional $\phi^4$ double-well model, we show that the leading deformation, which is just a squeeze, already cancels the one-loop divergence in the energy density of the domain wall soliton.  

% \vfill
%
% \end{titlepage}
\setcounter{footnote}{0}
\renewcommand{\thefootnote}{\arabic{footnote}}

\ifthenelse{\equal{\pr}{1}}
{
\maketitle
}{}

\section{Introduction}
Solutions of the linearized equations of motion of a classical field theory, which are superpositions of plane waves, correspond to the perturbative Fock space of multimeson states in the corresponding quantum field theory.  In Ref.~\cite{skyrme}, Skyrme suggested that even intrinsically nonlinear solutions of a classical field theory, which we will loosely call solitons, correspond to states in a quantum field theory.  Skyrme's suggestion was eventually transformed into a series of concrete formalisms \cite{dhn2,gs74,cl75,tom75,fk77,cq1,cq2,gw22} for treating solitons in quantum field theories.  

It is generally believed \cite{vinc72,cornwall74} that solitons correspond to approximately coherent states in quantum field theories.  Finding the soliton states and their quantum corrections is important for various applications.  In the case of perturbative excitations, one is usually not interested in quantum corrections to states because the LSZ formula allows one to calculate the S-matrix using only the uncorrected states. However no LSZ formula is known for scattering in a soliton sector.  In the case of perturbative states the spectrum can be calculated using standard interaction picture perturbation theory.  But in the case of soliton states, this is complicated by the zero modes, which imply that normal ordered products of operators have nonvanishing and in general divergent expectation values on the soliton ground state.  

Systematic approaches to finding quantum corrections to these coherent states were first considered at order $O(1)$ in Ref.~\cite{taylor78}, at one loop in Ref.~\cite{mekink} and beyond in Ref.~\cite{me2loop}.  Recently it has been appreciated that these quantum corrections have important consequences.  For example they may drastically increase the lifetime of the oscillon \cite{noiosc} and they also eliminate some divergences \cite{cocorr23}.  The present work concerns this last point.

In Ref.~\cite{erice}, Coleman noted that the coherent state construction, although it works wonderfully in 1+1 dimensions, leads to an infinite energy density in higher dimensions.  He called upon the students at the Erice school to find the correct construction of soliton states in higher dimensions.  In the present paper we answer this call, solving the simplest manifestation of this problem.  

In the (3+1)-dimensional $\phi^4$ double-well theory, the coherent state domain wall tension is divergent already at one loop.  The leading quantum correction to the coherent states was already implicit in Ref.~\cite{cahill76}.  In the true vacuum $\ovac$ all perturbative excitations must be in their ground state, so that $A_p\ovac_0=0$ where $A_p$ is the meson annihilation operator which creates plane waves and $\ovac_0$ is the leading order approximation to the true vacuum.  However in the soliton ground state, the perturbative excitations are not plane waves but rather are normal modes.  Thus the ground state must be annihilated by the operators $B_k$ that annihilate normal modes, as well as various momentum operators for zero modes.  These are related to the $A_p$ by a Bogoliubov transformation \cite{wentzel}.  A Bogoliubov transformation is implemented by a squeeze operator.  And so it is known that soliton states are not ordinary coherent states, but rather are squeezed coherent states.  The squeeze is the leading order correction to the coherent state, and all other corrections are suppressed by powers of the coupling.

Our main result is that the squeeze itself is sufficient to remove the one-loop divergence in the tension of the domain wall in the $(3+1)-$dimensional double-well model.  We announced this result already in Ref.~\cite{noi4dlett}.  However, a full derivation was not presented.  The present paper handles several important issues not treated there.  For example, the original theory is normal ordered at the bare mass scale, but this needs to be transformed to a finite mass scale.  The change in renormalization condition leads to new divergences.  We show in the present paper that these divergences only appear beyond one loop.  Furthermore, in Ref.~\cite{noi4dlett} we did not define the renormalization conditions, but rather simply wrote down the divergent parts of the mass and coupling counterterms that we claimed would lead to finite quantities.  In the present paper, renormalization conditions are chosen and imposed order by order in the vacuum sector.  

This computation is performed in the Schrodinger picture, which makes it far more complicated.  However, as stressed in Ref.~\cite{fk77}, calculations in the vacuum sector that are equal to those in the soliton sector in the ultraviolet imply that the cancellation of ultraviolet divergences in the vacuum sector is equivalent to the cancellation of those divergences in the soliton sector.  Thus, the seemingly masochistic calculations below in fact bring the added value of ensuring the finiteness of the corresponding quantities in the domain wall sector.

We also show that infrared divergences cancel, which was not shown in Ref.~\cite{noi4dlett}.  This cancellation is particularly delicate as even a small mismatch, when integrated over all of space in the vacuum sector, would lead to an infinite energy.

We begin in Sec.~\ref{teorsez} by reviewing the double-well model and its spontaneous symmetry breaking.  In Sec.~\ref{vacsez} we renormalize the ultraviolet divergences that arise at one loop by introducing counterterms for the overall energy, the mass and the coupling constant.  Wave function renormalization, which makes this theory trivial, is only necessary to remove the divergences at two loops and so will not appear here.  Once the counterterms are fixed, the renormalized mass and coupling are fixed, and our freedom is more or less exhausted, except for a choice of displacement operator to construct the coherent state\footnote{Even here, subleading corrections to the displacement operator will be compensated by the perturbative corrections to the state, and so there is no true freedom.  In fact, in Ref.~\cite{noi4dlett} the displacement operator was constructed using the bare parameters and here it is constructed using the renormalized parameters and, nonetheless, our results are consistent.}.  We therefore proceed to calculate the one-loop correction to the mass of the domain wall soliton in Sec.~\ref{wallsez}.

\section{The Theory} \label{teorsez}

\subsection{Definitions}

We will be interested in a theory of a scalar field $\phi(\vx)$ and its conjugate momentum $\pi(\vx)$ in 3+1 dimensions.  It is described by the Hamiltonian
\beq
\hat H=\int d^3\vx:\hat\ch^0(\vx):_{m_0}\hsp
\hat\ch^0(\vx)=\frac{\pi^2(\vx)+\sum_{i=1}^3\left(\partial_i \phi(\vx)\right)^2}{2}+\frac{\lambda_0 \phi^4(\vx)-m_0^2 \phi^2(\vx)}{4}+A
\eeq
where the normal ordering $::_{m_0}$ is defined at the mass  $m_0$.  More precisely, one uses the Schrodinger picture decomposition
\bea
\phi(\vx)&=&\pinv{3}{p}e^{-i\vp\cdot\vx}\left(A^{(0)\ddag}_\vp+\frac{A^{(0)}_{-\vp}}{2\omega^0_\vp}\right)\hsp 
\pi(\vx)=i\pinv{3}{p}e^{-i\vp\cdot\vx}\left(\omega^0_\vp A^{(0)\ddag}_\vp-\frac{A^{(0)}_{-\vp}}{2}\right)\nonumber\\
\omega^0_{\vp}&=&\sqrt{m_0^2+p^2}\hsp A^{(0)\ddag}_\vp=\frac{A^{(0)\dag}_\vp}{2\omega^0_\vp}
\eea
and places all $A^{(0)\ddag}$ to the left of the $A^{(0)}$.  The $c$-number $A$ is a counterterm that will be fixed momentarily.

Now we will introduce renormalized quantities $m$ and $\lambda$ via the definitions
\beq
m^2=m_0^2+\delta m^2\hsp \sl=\sqrt{\lambda_0}+\delta\sl \label{ct}
\eeq
where the counterterms $\delta m^2$ and $\delta\sl$ are taken to be of order $O(\lambda)$ and $O(\lambda^{3/2})$.  In principle, these may contain higher order contributions.  However this order is sufficient for the calculations in the present note.  We will also expand the counterterm $A$ in powers of $\lambda$
\beq
A=\sum_{i=0}^\infty A_i
\eeq
where $A_i$ is of order $O(\lambda^{i/2-1})$.

It is known that this theory is trivial in the sense that if $\lambda_0$ and $m_0$ are fixed, as one takes the ultraviolet cutoff $\Lambda\rightarrow\infty$, wave function renormalization makes the theory free \cite{froh82,ai21}.  While we are of course interested in the behavior as $\Lambda$ increases, we will always hold $\lambda$ and $m$ fixed while varying $\Lambda$.  This leads to a divergence in $\lambda_0$ and $m_0$ in the strict $\Lambda\rightarrow\infty$ limit.  Be this as it may, we are only interested in the question of whether the domain wall tension is bounded as $\Lambda$ increases with $\lambda$ and $m$ fixed.

\subsection{Changing the Normal Ordering}

Instead of normal ordering at the scale $m_0$, it will be convenient to normal order at the scale $m$, which remains bounded as the ultraviolet cutoff $\Lambda$ is taken to infinity.  Now, one uses the Schrodinger picture decomposition
\bea
\phi(\vx)&=&\pinv{3}{p}e^{-i\vp\cdot\vx}\left(A^{\ddag}_\vp+\frac{A_{-\vp}}{2\omega_\vp}\right)\hsp 
\pi(\vx)=i\pinv{3}{p}e^{-i\vp\cdot\vx}\left(\omega_\vp A^{\ddag}_\vp-\frac{A^{}_{-\vp}}{2}\right)\nonumber\\
\omega_{\vp}&=&\sqrt{m^2+p^2}\hsp A^{\ddag}_\vp=\frac{A^{\dag}_\vp}{2\omega_\vp}
\eea
and places all $A^{\ddag}$ to the left of the $A^{}$.  The corresponding Hamiltonian density is defined by
\beq
\hat H=\int d^3\vx:\hat\ch(\vx):_{m}.
\eeq

To evaluate $\hat\ch(\vx)$, we use the identities
\bea
 :\pi^2(\vx):_{m_0}&=&:\pi^2(\vx):_{m}+\frac{1}{2}\pinv{3}{p}\left({}{\omega_p}-{}{\omega^0_p}\right)\\
 \sum_j:\partial_j\phi(\vx)\partial_j\phi(\vx):_{m_0}&=& \sum_j:\partial_j\phi(\vx)\partial_j\phi(\vx):_{m}+\frac{1}{2}\pinv{3}{p}\left(\frac{p^2}{\omega_p}-\frac{p^2}{\omega^0_p}\right)\nonumber\\
:\phi^4(\vx):_{m_0}&=&:\phi^4(\vx):_{m}+6I:\phi^2(\vx):_m+3I^2\hsp :\phi^2(\vx):_{m_0}=:\phi^2(\vx):_{m}+I\nonumber\\
I&=&\frac{1}{2}\pinv{3}{p}\left(\frac{1}{\omega_p}- \frac{1}{\omega^0_p}\right)%\hsp\omega_p=\sqrt{m^2+p^2}%\hsp \omega^0_p=\sqrt{m_0^2+p^2}
\nonumber
\eea
to derive
\beq
\hat\ch(\vx)=\hat\ch^0(\vx)+\frac{3\lambda_0}{2}I\phi^2(\vx)+\frac{3\lambda_0}{4}I^2-\frac{3m_0^2}{4}I
+\frac{1}{4}\pinv{3}{p}\frac{\left(\omega_p-\omega^0_p\right)^2}{\omega_p}. \label{hh}
\eeq

Note that the momentum integrals are divergent.  They require ultraviolet cutoffs at a scale~$\Lambda$.  We always consider the large $\Lambda$ limit after the small $\lambda$ limit, so that it does not affect our power counting.  More precisely, we consider the limit such that $\lambda\Lambda^N$ goes  to zero for all $N$.

Let us count the powers of $\lambda$ in the new terms.  First, the $\lambda_0$ consists of $\lambda$ plus higher order terms, and so it is of at least order $O(\lambda)$.  Next, $I$ is an integral of an expression that is proportional to $\delta m^2$ plus higher powers of $\delta m^2$, and all terms are at least of the order of $\delta m^2$, which is of order $O(\lambda)$.  Finally, the expression $\sqrt{m^2+p^2}-\sqrt{m_0^2+p^2}$ again consists of powers of $\delta m^2$, of which each is of order at least $O(\lambda)$.  We thus conclude that the fifth term on the right hand side of Eq.~(\ref{hh}) is of order at least $O(\lambda^2)$, while the third begins at order $O(\lambda^3)$.  In the one loop treatment in this paper, we will not need any terms of order greater than $O(\lambda^{3/2})$, corresponding to $\delta\sl$.  Therefore we may and will simply drop these terms in the rest of this note.   %The fourth term is of order $O(\lambda)$, however it is a $c$-number and so will be compensated by the counterterm $A$.  

What about the second term?  Its coefficient is of order $O(\lambda^2)$, however it depends on $\phi(\vx)$ whose order is $O(1/\sl)$ as a result of the spontaneous symmetry breaking.  Thus the second term is of order $O(\lambda)$ overall and we need to keep it.  We also need to keep the fourth term, as it is of order $O(\lambda).$

Inserting (\ref{ct}) into the Hamiltonian density, we may now write it in terms of renormalized quantities and counterterms
\bea
\hat\ch(\vx)&=&\frac{\pi^2(\vx)+\sum_{i=1}^3\left(\partial_i \phi(\vx)\right)^2}{2}+\frac{\lambda \phi^4(\vx)-m^2 \phi^2(\vx)}{4}-\frac{3m^2I}{4}\label{hhp}\\
&&\hspace{-1cm}+\frac{\left(-2\sl\delta\sl+(\delta\sl)^2 \right) \phi^4(\vx)+\left(\delta m^2+6\left(\sl-\delta\sl\right)^2 I\right) \phi^2(\vx)}{4}+A%+\frac{3\delta m^2 I}{4}
\nonumber
\eea
where the second line contains the counterterms.
%\gre{9.1:we seem can drop the last term in (2.9) because when $I~ o(\lambda^1),\delta m~o(\lambda^1)$, so t is of order at least $\lambda^2$ }
\subsection{Spontaneous Symmetry Breaking}

This theory has two vacua, which are not located at $\langle\phi\rangle=0$.  This is inconvenient for perturbation theory, which is most easily written as an expansion in moments of $\phi(\vx)$.  However, we can shift $\langle\phi\rangle$ using the displacement operator
\beq
\dv={\rm{exp}}\left(-iv\dvx\ \pi(\vx)\right)
\eeq
which satisfies the relation
\beq
\phi(\vx)\dv=\dv\left(\phi(\vx)+v\right).
\eeq
We can use the unitary operator $\dv$ to transform the Hilbert space.  

We do not wish to change our theory, merely to redefine the fields and states so that one vacuum, which corresponds to the classical solution $\phi(x,t)=-m_0/\sqrt{2\lambda_0}$, now lies at the origin.  

As usual, such a passive transformation requires one to transform all operators as well.   More precisely, our passive transformation is defined as follows:  We act on all states with the operator $\dv$ and we conjugate all operators by $\dv$.  Therefore, in the new frame, time evolution is generated by the Hamiltonian density
\beq
H[\phi(\vx),\pi(\vx)]=\dv^\dag \hat H[\phi(\vx),\pi(\vx)]\dv=\hat H[\phi(\vx)+v,\pi(\vx)].
\eeq

Note that conjugation with $\dv$ only shifts fields by $c$-numbers and so does not affect the normal ordering.  Therefore, one may transform $\hat\ch$ to $\ch$ by simply replacing $\phi(\vx)$ with $\phi(\vx)+v$
\bea
\ch(\vx)&=&\frac{\pi^2(\vx)+\sum_{i=1}^3\left(\partial_i \phi(\vx)\right)^2}{2}+\frac{\lambda (\phi(\vx)+v)^4-m^2 (\phi(\vx)+v)^2}{4}\label{chsb}\\
&&+\frac{\left(-2\sl\delta\sl+(\delta\sl)^2 \right) (\phi(\vx)+v)^4}{4}\nonumber\\
&&+\frac{\left(\delta m^2+6\left(\sl-\delta\sl\right)^2 I\right)  (\phi(\vx)+v)^2}{4}-\frac{3%(
m^2%-\delta m^2)
I}{4}+A.\nonumber
\eea

Similarly to the counterterms, we will expand $v$ as

\beq
v=-\frac{m}{\sqrt{2\lambda}}+\delta v\hsp \delta v=\sum_{i=1}^\infty \delta v_i
\eeq
where $\delta v_i$ is of order $O(\lambda^{i/2})$.  Therefore our Hamiltonian density is
\bea
\ch(\vx)&=&\frac{\pi^2(\vx)+\sum_{i=1}^3\left(\partial_i \phi(\vx)\right)^2}{2}+\frac{\lambda \left(\phi(\vx)-\frac{m}{\sqrt{2\lambda}}+\delta v\right)^4-m^2 \left(\phi(\vx)-\frac{m}{\sqrt{2\lambda}}+\delta v\right)^2}{4}\nonumber\\
&&+\frac{\left(-2\sl\delta\sl+(\delta\sl)^2 \right) \left(\phi(\vx)-\frac{m}{\sqrt{2\lambda}}+\delta v\right)^4}{4}\nonumber\\
&&+\frac{\left(\delta m^2+6\left(\sl-\delta\sl\right)^2 I\right)  \left(\phi(\vx)-\frac{m}{\sqrt{2\lambda}}+\delta v\right)^2}{4}+\frac{-3%(
m^2%+\delta m^2)
I}{4}+A.\label{hpad}%\nonumber
\eea
Notice that in the last line the $3m^2I/4$ in the first term cancels a $-3m^2I/4$ in the second term.  These terms arose from the fact that we shifted the normal ordering when expanding about a maximum in the potential, and would never have appeared had we first expanded about the true vacuum and then changed the normal ordering.

\subsection{Expanding the Hamiltonian}

We will expand the Hamiltonian and Hamiltonian density in powers of $\lambda$
\beq
H=\sum H_i\hsp H_i=\dvx :\ch_i(\vx):_m
\eeq
where $H_i$ and $\ch_i(\vx)$ are of order $O(\lambda^{i/2-1})$.  We will set $A_1$ to zero, and we will check that this choice is consistent with our renormalization conditions at each order.

At leading order we find
\beq
\ch_0(\vx)=-\frac{m^4}{16\lambda}+A_0. \label{ch0}
\eeq
Next, a potentially dangerous  tadpole arises at nonperturbative order $O(1/\sl)$.  However, with the choice $A_1=0$, %the coefficient is exactly zero and so we learn 
$\ch_1(\vx)=0$.  The last order that is not perturbatively suppressed by powers of $\sl$ is
\beq
\ch_2(\vx)=\frac{\pi^2(\vx)+\sum_{i=1}^3\left(\partial_i \phi(\vx)\right)^2+m^2\phi^2(\vx)}{2}+\frac{m^4}{8\lambda}\left(-\frac{\delta\sl}{\sl}+\frac{\delta m^2}{m^2}\right)+A_2.
\eeq

The lowest order interaction term is
\beq
\ch_3(\vx)=-m\sqrt{\frac{\lambda}{2}}\phi^3(\vx)+\phi(\vx)\left[m^2\delta v_1+\frac{m^3}{\sqrt{2\lambda}}\left(\frac{\delta\sl}{\sqrt{\lambda}}-\frac{\delta m^2}{2m^2}\right)\right]+A_3.
\eeq
Note that the $\delta\sl$ and $\delta m^2$ diverge as the cutoff $\Lambda$ tends to infinity.  This divergence is not necessarily a problem, as the counterterms, like the bare terms, cannot be measured.  However, if $\delta v_1$ diverges, then the two vacuum sectors are infinitely separated and one may wonder whether the domain walls that we are studying exist.  It will be reassuring later that we will find that $\delta v_1$ in fact remains bounded as $\Lambda$ tends to infinity.

The next to leading order interaction is
\bea
\ch_4(\vx)&=&\frac{\lambda}{4}\phi^4(\vx)+\left[-\frac{3m\sl\delta v_1}{\sqrt{2}}-\frac{3m^2\delta\sl}{2\sl}+\frac{\delta m^2}{4} \right]\phi^2(\vx)\\
&&\hspace{-1cm}+{m^2\delta v_2}{}\phi(\vx)+\frac{m^4\left(\delta\sl\right)^2}{16\lambda^2}+\frac{m^3\delta v_1}{\sqrt{2\lambda}}\left[ 
\frac{\sl}{\sqrt{2}m}\delta v_1+\frac{\delta\sl}{\sl}-\frac{\delta m^2}{2m^2}\right]+A_4.\nonumber
\eea
The highest order interaction that we will need is
\bea
\ch_5(\vx)&=&\left[\lambda\delta v_1+\sqrt{2}m\delta\sl\right]\phi^3(\vx)+\left[-\frac{3m\sl\delta v_2}{\sqrt{2}} \right]\phi^2(\vx)\\
&&\hspace{-1.5cm}+\left[m^2\delta v_3-3m\sqrt{\frac{\lambda}{2}}\left(\delta v_1\right)^2-3\frac{m^2\delta\sl\delta v_1}{\sl} -\frac{m^3\left(\delta\sl\right)^2}{2\sqrt{2}\lambda^{3/2}}+\frac{\delta m^2\delta v_1}{2}-\frac{3m\sl I}{\sqrt{2}}\right]\phi(\vx)\nonumber\\
&&\hspace{-1.5cm}+\frac{m^3\delta v_2}{\sqrt{2\lambda}}\left[\frac{\sqrt{2\lambda}}{m}\delta v_1 +\frac{\delta\sl}{\sl}
-\frac{\delta m^2}{2m^2}\right]+A_5.\nonumber
\eea

\section{Renormalizing the Vacuum Sector} \label{vacsez}

\subsection{Renormalization Conditions} \label{rcsez}

The Hamiltonian $\hat{H}$ contained three counterterms: $A$, $\delta m^2$ and $\delta\sl$.  The shift to $H$ also depended on a free parameter $\delta v$.  To fix these four quantities, we need four renormalization conditions.  We choose the following
\begin{enumerate}
  \item \hypertarget{first}{$H\ovac=0$}
  \item \hypertarget{second}{$H|\vp\rangle=\ovp{}|\vp\rangle$} 
  \item \hypertarget{third}{${}_0\langle \vp_1 \vp_2|\vp_0\rangle_{i\neq 1}=0$}%\gre{why not$_0\langle \vp_1 \vp_2|\vp_0\rangle_{i\neq 0}=0$, each order should satisfied} \red{This is a key point.  The idea of this renormalization condition is that all of the subleading corrections to the 3 meson vertex should vanish, so that the interaction is given by $m\sl$.  The leading vertex however is of order $O(\sl)$ and so it corresponds to $i=1$. The $i=0$ term is easy to calculate and you can see that it is zero.}\gre{Thank you I see it}
  \item \hypertarget{fourth}{$\langle\Omega|\phi(\vx)|\Omega\rangle=0$}.
\end{enumerate}
Here $\ovac$ is the vacuum. $|\vp_1\cdots \vp_n\rangle$ is an $n$-meson state which is simultaneously a Hamiltonian eigenstate and also a momentum eigenstate with momentum $\sum \vp_i$.  Each state is expanded
\beq
|\Psi\rangle=\sum_{i=0}^\infty |\Psi\rangle_i
\eeq
where $|\Psi\rangle_i$ is of order $O(\lambda^{i/2})$.  We define $\ovac$ to be the Hamiltonian eigenstate such that
\beq
A_\vp\ovac_0=0.
\eeq
The states are normalized by the three conditions
\renewcommand{\labelenumi}{\Roman{enumi})}
\begin{enumerate}
\item ${}_0\langle \Omega|\Omega\rangle_0=1$ 
  \item $|\vp_1\cdots \vp_n\rangle_0=A^\ddag_{\vp_1}\cdots A^\ddag_{\vp_n}|\Omega\rangle_0$  
  \item 
   \hypertarget{terzo}{${}_0\langle \vp_1\cdots \vp_m|\vp_1\cdots \vp_m\rangle_{n>0}=0.$}
\end{enumerate}
However the renormalization conditions are independent of the normalization.

Note that the renormalization conditions are expressed in terms of states in the vacuum sector Fock space.  Intuitively, we demand a cancellation of divergences in the vacuum sector.  Once this is done, the four parameters are fixed.  The Hamiltonian eigenstates in the domain wall sector are then determined, and there is no more freedom that can be used to eliminate any domain wall sector divergences.

This Schrodinger picture renormalization is far less efficient then the usual interaction picture renormalization.  We use it because the Schrodinger picture calculations are identical in the ultraviolet to corresponding calculations in the domain wall sector.  The general argument of Ref.~\cite{fk77} then implies that ultraviolet finiteness in the vacuum sector of a given quantity will lead to finiteness in the domain wall sector of the same quantity.

\subsection{Order $O(1)$ and below}

At order $O(1/\lambda)$ we consider $\ch_0$ given in Eq.~(\ref{ch0}).  The \hyperlink{first}{first} renormalization condition implies
\beq
A_0=\frac{m^4}{16\lambda}\hsp H_0=0. \label{a0}
\eeq
As $\ch_1$ vanishes, the renormalization conditions are trivially satisfied at order $O(1/\sl)$.

To proceed to order $O(1)$, note that the free Hamiltonian may be written
\beq
H_2=\pinv{3}{p}\omega_{\vp}A^\ddag_\vp A_\vp+\dvx\left[ \frac{m^4}{8\lambda}\left(-\frac{\delta\sl}{\sl}+\frac{\delta m^2}{m^2}\right)+A_2\right].
\eeq
The \hyperlink{first}{first} renormalization condition at $O(1)$ then implies
\beq
A_2= \frac{m^4}{8\lambda}\left(\frac{\delta\sl}{\sl}-\frac{\delta m^2}{m^2}\right)\hsp H_2=\pinv{3}{p}\omega_{\vp}A^\ddag_\vp A_\vp. \label{a2}
\eeq
This automatically also satisfies the \hyperlink{second}{second} renormalization condition at order $O(1)$.  The \hyperlink{third}{third} renormalization condition is trivially satisfied at this order as
\beq
{}_0\langle \vp_1 \vp_2|\vp_3\rangle_0=4\ovp 1\ovp 2 {}_0\langle\Omega|A_{\vp_1}A_{\vp_2} A^{\ddag}_{\vp_3}\ovac_0=4\ovp 1\ovp 2 {}_0\langle\Omega|[A_{\vp_1}A_{\vp_2}, A^{\ddag}_{\vp_3}]\ovac_0=0
\eeq
which vanishes as
\beq
[A_{\vp_1}, A^{\ddag}_{\vp_2}]=(2\pi)^3\delta^3(\vp_1-\vp_2)
\eeq
and $A\ovac_0=0$.  Similarly the \hyperlink{fourth}{fourth} renormalization condition is satisfied at this order
\beq
{}_0\langle\Omega|\phi(\vx)\ovac_0=\pinv{3}{p}e^{-i\vp\cdot\vx}\left({}_0\langle\Omega|A^{\ddag}_\vp\ovac_0+\frac{{}_0\langle\Omega|A_{-\vp}\ovac_0}{2\omega_\vp}\right)=0.
\eeq

\subsection{Order $O(\sl)$}

The vacuum sector can be decomposed into $n$-meson Fock spaces, which are spanned respectively by the states
\beq
|\vp_1\cdots \vp_n\rangle_0=A^\ddag_{\vp_1}\cdots A^\ddag_{\vp_n}\ovac_0.
\eeq
The operator $H_2$ preserves $n$.  For any given state $|\psi\rangle$ in the vacuum sector, we will write this decomposition as
\beq
|\psi\rangle=\sum_n |\psi\rangle^n
\eeq
where $|\psi\rangle^n$ lies in the $n$-meson Fock space, and so is a linear combination of the states $|\vp_1\cdots \vp_n\rangle_0$.

At order $O(\sl)$ the \hyperlink{first}{first} renormalization condition is
\beq
H_2\ovac_1=-H_3\ovac_0. \label{slcond}
\eeq
Let us impose this condition in each $n$-meson Fock space.

First, note that the zero-meson Fock space consists of $\ovac_0$ which is annihilated by $H_2$.  Therefore, restricting to the zero-meson Fock space, the left hand side vanishes, and so the right hand side must also vanish.  The right hand side, restricted to the zero-meson Fock space, is $A_3\ovac_0$ and so we learn
\beq
A_3=0.
\eeq

\begin{figure}[htbp]
\centering
\includegraphics[width = 0.35\textwidth]{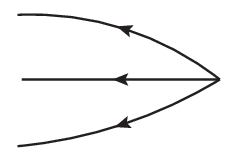}
\caption{This graph represents the calculation of $\ovac_1^3$, which is $-H_2^{-1}H_3\ovac_0^0$.  The vertex represents the operator $-H_2^{-1}H_3$.    The order $i$ of $\ovac_i$ increases as one moves to the left.  Consider a vertical slice.  It intersects some number of lines $n$.  This represents the $n$-meson Fock space.  To the right of the vertex there are no lines, reflecting the fact that $\ovac_0=\ovac_0^0$ lives in the zero-meson Fock space.  To the left, there are three lines, as we are calculating a contribution $\ovac_1^3$ to $\ovac_1$ in the three-meson Fock space.} \label{v1fig}
\end{figure}

To shorten expressions, let us introduce the notation
\beq
\delta v\p_1=\delta v_1+\frac{m}{\sqrt{2\lambda}}\left(\frac{\delta\sl}{\sqrt{\lambda}}-\frac{\delta m^2}{2m^2}\right). 
\eeq
Then
\beq
H_3\ovac_0=-m\sqrt{\frac{\lambda}{2}}\pinv{3}{p_1}\pinv{3}{p_2}|\vp_1,\vp_2,-\vp_1-\vp_2\rangle_0+m^2\delta v\p_1 |\vp=\Vec{0}\rangle_0.
\eeq
\par 
Now we can solve (\ref{slcond}) by inverting $H_2$, where the inverse is unique once normalization condition (\hyperlink{terzo}{III}) in Subsec.~\ref{rcsez} is imposed
\beq
\ovac_1=m\sqrt{\frac{\lambda}{2}}\pinv{3}{p_1}\pinv{3}{p_2}\frac{|\vp_1,\vp_2,-\vp_1-\vp_2\rangle_0}{\ovp 1+\ovp 2+\omega_{\vp_1+\vp_2}}-m\delta v\p_1 |\vp=\Vec{0}\rangle_0. \label{vac1}
\eeq
With this perturbative correction to $\ovac$, the \hyperlink{first}{first} renormalization condition is solved.

We will introduce a graphical representation of such quantum corrections to states, which will be helpful later when we want to specify which contributions we are calculating.  In the case of Eq.~(\ref{vac1}), the graphical representation is in Fig.~\ref{v1fig}.  The perturbative corrections $|\psi\rangle_i$ are calculated with $i$ increasing as one moves along the arrows, which we will point to the left.  Each line represents a meson and each vertex represents $(E-H_2)^{-1}H_j$ for some $j$.  Therefore, if there are $n$ incoming mesons and $m$ outgoing mesons, the graph describes a contribution to $|\vp_1\cdots \vp_n\rangle^m$.

Let us skip to the fourth condition.  At order $O(\sl)$
\beq
\langle \Omega| \phi(\vx) \ovac=2\ {}_0\langle \Omega| \phi(\vx) \ovac_1=-\delta v\p_1
\eeq
and so we learn
\beq
\delta v\p_1=0\hsp
\delta v_1=\frac{m}{\sqrt{2\lambda}}\left(-\frac{\delta\sl}{\sqrt{\lambda}}+\frac{\delta m^2}{2m^2}\right)\hsp \ch_3(\vx)=-m\sqrt{\frac{\lambda}{2}}\phi^3(\vx). \label{v1}
\eeq

Now we are ready for the \hyperlink{second}{second} renormalization condition.  To make the calculation faster, we will decompose the vacuum sector states by meson number $n$
\beq
|\Psi\rangle=\sum_n |\Psi\rangle^n\hsp |\Psi\rangle_i=\sum_n |\Psi\rangle_i^n
\eeq
which is defined as the number of $A^\ddag$ operators that act on $\ovac_0$.  We will also decompose terms in the Hamiltonian by meson number, defined so that the total meson number of an operator plus a state is conserved when an operator acts on a state.  In the case of the leading interaction, this decomposition is
\bea
H_3&=&\sum_{n=0}^3 H_3^{3-2n}\hsp
H_3^{3}=-\frac{m\sl}{\sqrt{2}}\pinv{3}{p_1}\pinv{3}{p_2} \Ad 1\Ad 2 A^\ddag_{-\vp_1-\vp_2}\\
H_3^{1}&=&-\frac{3m\sl}{2\sqrt{2}}\pinv{3}{p_1}\pinv{3}{p_2} \frac{\Ad 1\Ad 2 A_{\vp_1+\vp_2}}{\omega_{\vp_1+\vp_2}}\nonumber\\
H_3^{-1}&=&-\frac{3m\sl}{4\sqrt{2}}\pinv{3}{p_1}\pinv{3}{p_2} \frac{\Ad 1 A_{-\vp_2} A_{\vp_1+\vp_2}}{\ovp 2\omega_{\vp_1+\vp_2}}\nonumber\\
H_3^{-3}&=&-\frac{m\sl}{8\sqrt{2}}\pinv{3}{p_1}\pinv{3}{p_2} \frac{A_{-\vp_1} A_{-\vp_2} A_{\vp_1+\vp_2}}{\ovp 1\ovp 2\omega_{\vp_1+\vp_2}}.
\eea

\begin{figure}[htbp]
\centering
\includegraphics[width = 0.6\textwidth]{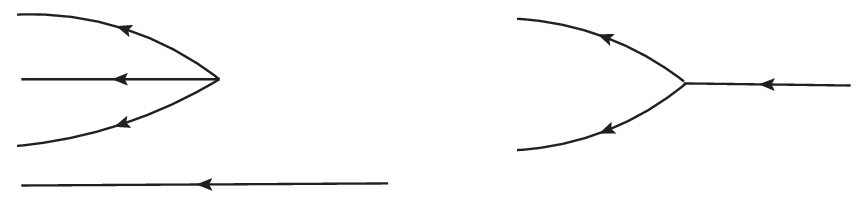}
\caption{This is a graphical representation of $|\vp\rangle_1^4$ (left) and $|\vp\rangle_1^2$ (right).  In each case, to the right of the vertex there is a single meson, as $|\vp\rangle_0=|\vp\rangle_0^1$ is contained in the one-meson Fock space.}\label{m1fig}
\end{figure}

Now we decompose the \hyperlink{second}{second} renormalization condition at meson number $n$
\beq
H_3^{n-1}|\vp\rangle^1_0=(\ovp{}-H_2)|\vp\rangle^n_1.
\eeq
As $|\vp\rangle_0=A^\ddag_\vp\ovac_0$, it contains a single meson and so will be annihilated by $H_3^{-1}$ as a result of the normal ordering.  Thus the left hand side is only nonvanishing at meson numbers $n=2$ and $n=4$.  Remembering that $H_3^{n-1}$ has $2-n/2$ annihilation operators and $|\vp\rangle_0$ contains one creation operator, the commutator yields a factor of the number of contractions, which in this case is unity for both $n=2$, which has one contraction, and $n=4$ which has none.

Altogether we find
\bea
H_3^3|\vp\rangle^1_0&=&-\frac{m\sl}{\sqrt{2}}\pinv{3}{p_1}\pinv{3}{p_2}|\vp,\vp_1,\vp_2,-\vp_1-\vp_2\rangle_0\\
H_3^1|\vp\rangle^1_0&=&-\frac{3m\sl}{2\sqrt{2}\ovp{}}\pinv{3}{p_1}|\vp_1,\vp-\vp_1\rangle_0.\nonumber
\eea
Inverting the $(\ovp{}-H_2)$ one finds
\bea
|\vp\rangle_1^4&=&\frac{m\sl}{\sqrt{2}}\pinv{3}{p_1}\pinv{3}{p_2}\frac{|\vp,\vp_1,\vp_2,-\vp_1-\vp_2\rangle_0}{\ovp 1 +\ovp 2 +\omega_{\vp_1+\vp_2}}\\
|\vp\rangle^2_1&=&\frac{3m\sl}{2\sqrt{2}\ovp{}}\pinv{3}{p_1}\frac{|\vp_1,\vp-\vp_1\rangle_0}{\ovp 1+\omega_{\vp-\vp_1}-\ovp{}}\nonumber
\eea
which is depicted in Fig.~\ref{m1fig}.

The \hyperlink{third}{third} renormalization condition at this order is the case $i=1$.  However the renormalization condition explicitly does not apply to $i=1$, and so it is trivially satisfied.

\subsection{Order $O(\lambda)$}

\subsubsection{The Vacuum State $\ovac$}

At order $O(\lambda)$ the \hyperlink{first}{first} renormalization condition restricted to the $n$-meson Fock space is
\beq
H_4^n\ovac^0_0+H_3^{n-3}\ovac_1^3=-H_2\ovac_2^n. \label{lam1}
\eeq

Let us define the shorthand,
\bea
A_4\p&=&\frac{m^4\left(\delta\sl\right)^2}{16\lambda^2}+\frac{m^3\delta v_1}{\sqrt{2\lambda}}\left[ 
\frac{\sl}{\sqrt{2}m}\delta v_1+\frac{\delta\sl}{\sl}-\frac{\delta m^2}{2m^2}\right]+A_4\nonumber\\
&=&\frac{m^4\left(\delta\sl\right)^2}{16\lambda^2}-\frac{m^2\delta v_1^2}{2}+A_4.%\hsp \delta v_2\p=\delta v_2+\frac{3I}{4m}.
\eea
%\blu{Not that it matters, but the RHS can be simplified to $+ \frac{m^{4}(\delta g)^{2}}{16g^{4}} - \frac{m^{2}\delta v_ {1}^{2}}{2} + A_ 4$.}
Then $\ch_4(\vx)$ reduces to
\bea
\ch_4(\vx)&=&\frac{\lambda}{4}\phi^4(\vx)-\frac{\delta m^2}{2} \phi^2(\vx)+m^2\delta v_2\phi(\vx)+A_4\p.
\eea

Let us begin with the one-meson sector $n=1$.  As $H_3^{-2}=0$, on the left hand side we need only act with $\ch_4^1$ yielding
\beq
H_4^1\ovac_0=m^2\delta v_2 |\vp=\vec{0}\rangle_0.
\eeq
Inverting $H_2$ one finds
\beq
\ovac_2^1=-m\delta v_2 |\vp=\vec{0}\rangle_0.
\eeq
This leads to a tadpole
\beq
\langle \Omega| \phi(\vx) \ovac=2\ {}_0\langle \Omega| \phi(\vx) \ovac_2=-\delta v_2.
\eeq
The \hyperlink{fourth}{fourth} renormalization condition then implies
\beq
\delta v_2=0\hsp
%\delta v_2=-\frac{3I}{4m}\hsp
\ch_4(\vx)=\frac{\lambda}{4}\phi^4(\vx)-\frac{\delta m^2}{2} \phi^2(\vx)+A_4\p.
\eeq
The fact that $\delta v_2$ is finite implies that the distance between the vacua is finite, which is a consistency check of our calculation thus far.
%This divergence $\delta v_2$ is unphysical, as it implies that the two vacua become infinitely separated.  This problem can be solved by including an $O(\lambda^{3/2})$ contribution to $\delta m^2$ and a $O(\lambda^2)$ contribution to $\delta\sl$.  However, these higher order counterterms, and even $\delta v_2$ itself, do not affect the one-loop domain wall tension and so we will ignore them here, simply noting that the tadpole term vanishes as a result of the \hyperlink{third}{third} renormalization condition.  We will return to these terms in future work when we turn to the two-loop tension.  
%\red{$\delta v_2$ is divergent, meaning that the difference in VEVs between the vacua is divergent.  It is proportional to $I$, which is divergent because we normal ordered at a divergent scale $m_0$.  Is this divergent VEV a pathology?  Will it imply at infinite mass at some order, for example?  If so, maybe we need to define the original Hamiltonian normal ordered at some finite mass scale.  We can't make it $m$, because $m$ is something that we choose ourselves, not part of the definition of the theory.}

Now we are ready to decompose $H_4$ by meson number
\bea
H_4&=&\sum_{n=0}^4 H_4^{4-2n}\hsp
H_4^{4}=\frac{\lambda}{4}\pinv{3}{p_1}\pinv{3}{p_2}\pinv{3}{p_3} \Ad 1\Ad 2\Ad 3 A^\ddag_{-\vp_1-\vp_2-\vp_3}\\
H_4^{2}&=&\frac{\lambda}{2}\pinv{3}{p_1}\pinv{3}{p_2}\pinv{3}{p_3} \frac{\Ad 1\Ad 2\Ad 3 A_{\vp_1+\vp_2+\vp_3}}{\omega_{\vp_1+\vp_2+\vp_3}}-\frac{\delta m^2}{2}\pinv{3}{p}\Ad {}A^\ddag_{-\vp}\nonumber\\
H_4^{0}&=&\frac{3\lambda}{8}\pinv{3}{p_1}\pinv{3}{p_2}\pinv{3}{p_3} \frac{\Ad 1\Ad 2A_{-\vp_3} A_{\vp_1+\vp_2+\vp_3}}{\ovp 3\omega_{\vp_1+\vp_2+\vp_3}}-\frac{\delta m^2}{2}\pinv{3}{p}\frac{\Ad {}A_{\vp}}{\ovp{}}+\dvx A\p_4\nonumber\\
H_4^{-2}&=&\frac{\lambda}{8}\pinv{3}{p_1}\pinv{3}{p_2}\pinv{3}{p_3} \frac{\Ad 1 A_{-\vp_2}A_{-\vp_3} A_{\vp_1+\vp_2+\vp_3}}{\ovp 2\ovp 3\omega_{\vp_1+\vp_2+\vp_3}}-\frac{\delta m^2}{8}\pinv{3}{p}\frac{A_{-\vp}A_{\vp}}{\ovp{}^2}\nonumber\\
H_4^{-4}&=&\frac{\lambda}{64}\pinv{3}{p_1}\pinv{3}{p_2}\pinv{3}{p_3} \frac{A_{-\vp_1} A_{-\vp_2}A_{-\vp_3} A_{\vp_1+\vp_2+\vp_3}}{\ovp 1\ovp 2\ovp 3\omega_{\vp_1+\vp_2+\vp_3}}.\nonumber
\eea

\begin{figure}[htbp]
\centering
\includegraphics[width = 0.85\textwidth]{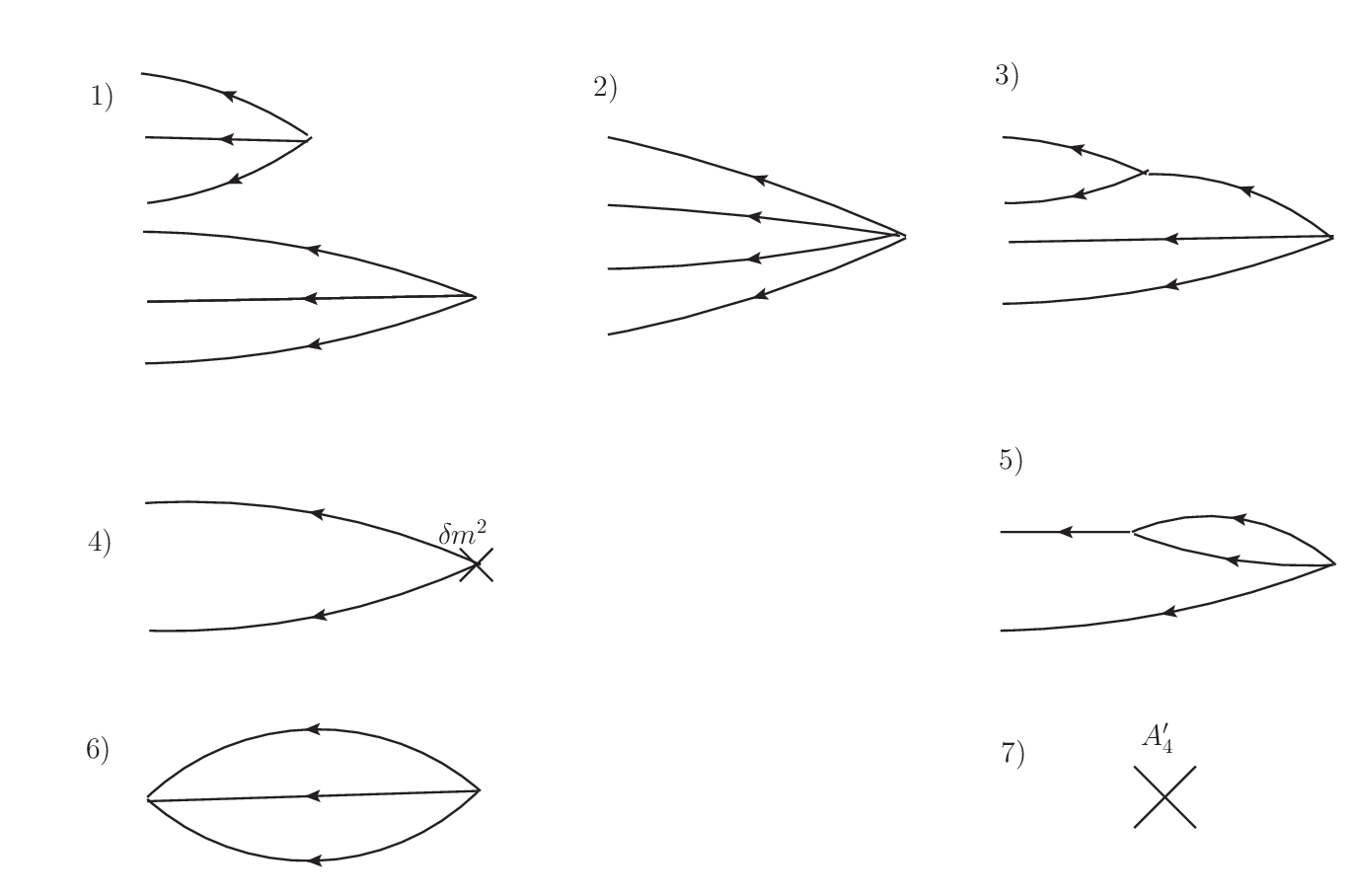}
\caption{We represent the contributions to $\ovac_2^6$ in panel (1), $\ovac_2^4$ in panels $(2)$ and $(3)$, $\ovac_2^2$ in panels $(4)$ and $(5)$ and $\ovac_2^0$ in panels $(6)$ and $(7)$.}\label{v2fig}
\end{figure}

We have now assembled all of the ingredients to solve the renormalization condition (\ref{lam1}) one meson number $n$ at a time, with the contributions drawn in Fig.~\ref{v2fig}.  At $n=6$, the only contribution to the left hand side arises from
\beq
H_3^3\ovac^3_1=-\frac{m^2\lambda}{2}\pinv{3}{p_1}\pinv{3}{p_2}\pinv{3}{p_3} \pinv{3}{p_4}\frac{|\vp_1,\vp_2,-\vp_1-\vp_2,\vp_3,\vp_4,-\vp_3-\vp_4\rangle_0}{\ovp 1+\ovp 2+\omega_{\vp_1+\vp_2}}.
\eeq
Inverting $H_2$ one then finds
\beq
\ovac_2^6=\frac{m^2\lambda}{2}\pinv{3}{p_1}\cdots \frac{d^3\vp_4}{(2\pi)^3}\frac{|\vp_1,\vp_2,-\vp_1-\vp_2,\vp_3,\vp_4,-\vp_3-\vp_4\rangle_0}{\left(\ovp 1+\ovp 2+\omega_{\vp_1+\vp_2}\right)\left(\omega_{\vp_1+\vp_2}+\omega_{\vp_3+\vp_4}+\sum_{i=1}^4 \ovp i
\right)}.
\eeq

At the lower meson numbers, both $H_3$ and $H_4$ contribute.  At meson number $n=4$ these contributions are
\bea
H_4^4\ovac_0&=&\frac{\lambda}{4}\pinv{3}{p_1}\pinv{3}{p_2}\pinv{3}{p_3}|\vp_1,\vp_2,\vp_3,-\vp_1-\vp_2-\vp_3\rangle_0\\
H_3^1\ovac_1^3&=&-\frac{9m^2\lambda}{4}\pinv{3}{p_1}\pinv{3}{p_2}\pinv{3}{p_3}\frac{|\vp_1,\vp_2,\vp_3,-\vp_1-\vp_2-\vp_3\rangle_0}{\omega_{\vp_1+\vp_2}\left(\ovp 1+\ovp 2 +\omega_{\vp_1+\vp_2}\right)}\nonumber
\eea

% \blu{
% Hi guys, I found the following result instead, it is not clear to me if my result and Jarah's are equivalent...
% \begin{align*}
% -H_ {2}^{-1} H_ {3}\left\lvert \Omega_ {1} \right\rangle =& \left( \frac{3mg}{2} \right)^{2} \int \frac{d^{3}p_ {1}d^{3}p_ {2}d^{3}p_ {3}}{(2\pi)^{9}} \;  \frac{1}{\omega_ {1+2}(\omega_ {1+2}+\omega_ {3}+\omega_ {1+2+3})} \\
% &\times \frac{\left\lvert \vec{p}_ {1,2,3,-1-2-3} \right\rangle_ {0}}{(\omega_ {1}+\omega_ {2}+\omega_ {3}+\omega_ {1+2+3})}.
% \end{align*}
% }

% \red{Your result looks like the first term in the square brackets of my expression.  The second term I think comes from $H_4^4$ ... but since you just calculated the $H_3$ term I think this is consistent.}

% \blu{Hi Jarah, yes this is just the $H_3$ part, but the difference comes from the denominator, I got $\frac{...}{...(\omega_{1+2}+\omega_{3}+\omega_{1+2+3})}$, while you have $\frac{...}{...(\omega_{1}+\omega_{2}+\omega_{1+2})}$. Are the the same after the integral?}

% \blu{Wait, I found a mistake in my code, let me redo the calculation...}

% \blu{yep I got the same result.}

% \red{There is a permutation symmetry here.  You can free exchange $1$, $2$, $3$ and $1+2+3$.  Also you can change the signs of the $k$'s.  I think that you can get from your result to mine with the substitutions $k_3\rightarrow -k_1$ and $k_1+k_2+k_3\rightarrow k_2$.  Then the first subscript $k_1+k_2=k_1+k_2+k_3-k_3$ goes to $k_2- - k_1=k_1+k_2$ so it does not change. Do you agree?}

% \blu{Yes indeed... I agree. I'll comment this part tomorrow.}

where we have included a factor of three in $H_3^1\ovac_1^3$ arising from the three possible contractions of the $A$ in $H_3^1$ with the three $A^\ddag$ operators in $\ovac_1^3$.  Inverting $H_2$ we find 
\beq
\ovac_2^4=\frac{\lambda}{4}\pinv{3}{p_1}\cdots \frac{d^3\vp_3}{(2\pi)^3}\left[\frac{9m^2}{\omega_{\vp_1+\vp_2}\left(\ovp 1+\ovp 2 +\omega_{\vp_1+\vp_2}\right)}-1 
\right]\frac{|\vp_1,\vp_2,\vp_3,-\vp_1-\vp_2-\vp_3\rangle_0}{\ovp 1+\ovp 2+\ovp 3+\omega_{\vp_1+\vp_2+\vp_3}}.
\eeq
The calculation at meson number $n=2$ is similar, with contributions
\bea
H_4^2\ovac_0&=&-\frac{\delta m^2}{2}\pinv{3}{p}|\vp,-\vp\rangle_0\\
H_3^{-1}\ovac_1^3&=&-\frac{9m^2\lambda}{4}\pinv{3}{p}\ppinv{3}{p}\frac{|\vp,-\vp\rangle_0}{\ovpp{}\omega_{\vp+\vpp}\left(\ovp{}+\ovpp{}+\omega_{\vp+\vpp} \right)}\nonumber
%\pinv{3}{p_1}\pinv{3}{p_2}\pinv{3}{p_3}\frac{|\vp_1,\vp_2,\vp_3,-\vp_1-\vp_2-\vp_3\rangle_0}{\omega_{\vp_1+\vp_2}\left(\ovp 1+\ovp 2 +\omega_{\vp_1+\vp_2}\right)}
\eea
where we have included a factor of six from the possible contractions of the two $A$ operators in $H_3^{-1}$ with the three $A^\ddag$ operators in $\ovac_1^3$.  Note that the $\vpp$ integral is logarithmically divergent.  Any logarithmic divergence in a state could take us out of the vacuum sector Fock space.  The corresponding component of the state is 
\beq
\ovac_2^2=\pinv{3}{p}\left[\frac{\delta m^2}{4}+\frac{9m^2\lambda}{8}\ppinv{3}{p}\frac{1}{\ovpp{}\omega_{\vp+\vpp}\left(\ovp{}+\ovpp{}+\omega_{\vp+\vpp} \right)}
\right]\frac{|\vp,-\vp\rangle_0}{\ovp{}}.
\eeq

% \blu{
% I got instead
% $$\left\lvert \Omega \right\rangle^{(2)}_ {2}=\int \frac{d^{3}p_ {1}}{(2\pi)^{3}} \, \left\lbrace \frac{\delta m^{2}}{4}+\frac{27m^{2}g^{2}}{8} \int \frac{d^{3}p_ {2}}{(2\pi)^{3}} \, \frac{1}{\omega_ {2}\omega_ {1+2}(\omega_ {2}+\omega_ {2}+\omega_ {1+2})}  \right\rbrace \frac{\left\lvert \vec{p}_ {1,-1} \right\rangle}{\omega_ {1}} .
% $$
% The second term differs by a factor of three, everything else is the same.
% }

We therefore need the expression in the square brackets to be finite.  This constrains $\delta m^2$ to be, up to a finite quantity
\beq
\delta m^2\sim -\frac{9m^2\lambda}{4}\pinv{3}{p}\frac{1}{p^3}+{\rm{finite}}\hsp p=|\vp|. \label{dma}
\eeq
Momentarily we will calculate $\delta m^2$ and check this condition.

Finally we turn to the zero-meson Fock space.  Here the relevant term in $H_4^0$ contains a $\dvx$ of an $\vx$-independent quantity and so is infrared divergent.  Similarly, working as above one would find that $H_3^{-3}\ovac_1^3$ contains a squared Dirac delta, and so it is also divergent.  The right hand side of (\ref{lam1}) vanishes as we have set $\ovac_2^0=0$.

The problem is that we may only perform the $\vx$ integration after summing these terms.  Before the $\vx$ integration, the two contributions are
\bea
\ch^0_4(\vx)\ovac_0&=&A\p_4\ovac_0\\
\ch_3^{-3}(\vx)\ovac_1^3&=&-\frac{3m^2\lambda}{8}\pinv{3}{p_1}\pinv{3}{p_2}\frac{\ovac_0}{\ovp 1\ovp 2\omega_{\vp_1+\vp_2}\left(\ovp1+\ovp2+\omega_{\vp_1+\vp_2}\right)}\nonumber
\eea
where we have included a factor of six in the second term for the $3!$ contractions of the three $A$ operators with the three $A^\ddag$ operators.  

The sum of these two terms needs to be exactly zero, because otherwise the $\vx$ integration will lead to a divergence.  Thus we conclude
\beq
A_4\p=\frac{3m^2\lambda}{8}\pinv{3}{p_1}\pinv{3}{p_2}\frac{1}{\ovp 1\ovp 2\omega_{\vp_1+\vp_2}\left(\ovp1+\ovp2+\omega_{\vp_1+\vp_2}\right)}. \label{a4}
\eeq
This exhibits a quadratic ultraviolet divergence.

\subsubsection{The One-Meson State $|\vp\rangle$}

The \hyperlink{second}{second} renormalization condition
\beq
H_4^{n-1}|\vp\rangle^1_0+H_3^{n-4}|\vp\rangle^4_1+H_3^{n-2}|\vp\rangle^2_1=(\ovp{}-H_2)|\vp\rangle_2^n \label{lamp}
\eeq
yields the subleading correction $|\vp\rangle_2$ to the one-meson state $|\vp\rangle$.

\begin{figure}[htbp]
\centering
\includegraphics[width = 0.85\textwidth]{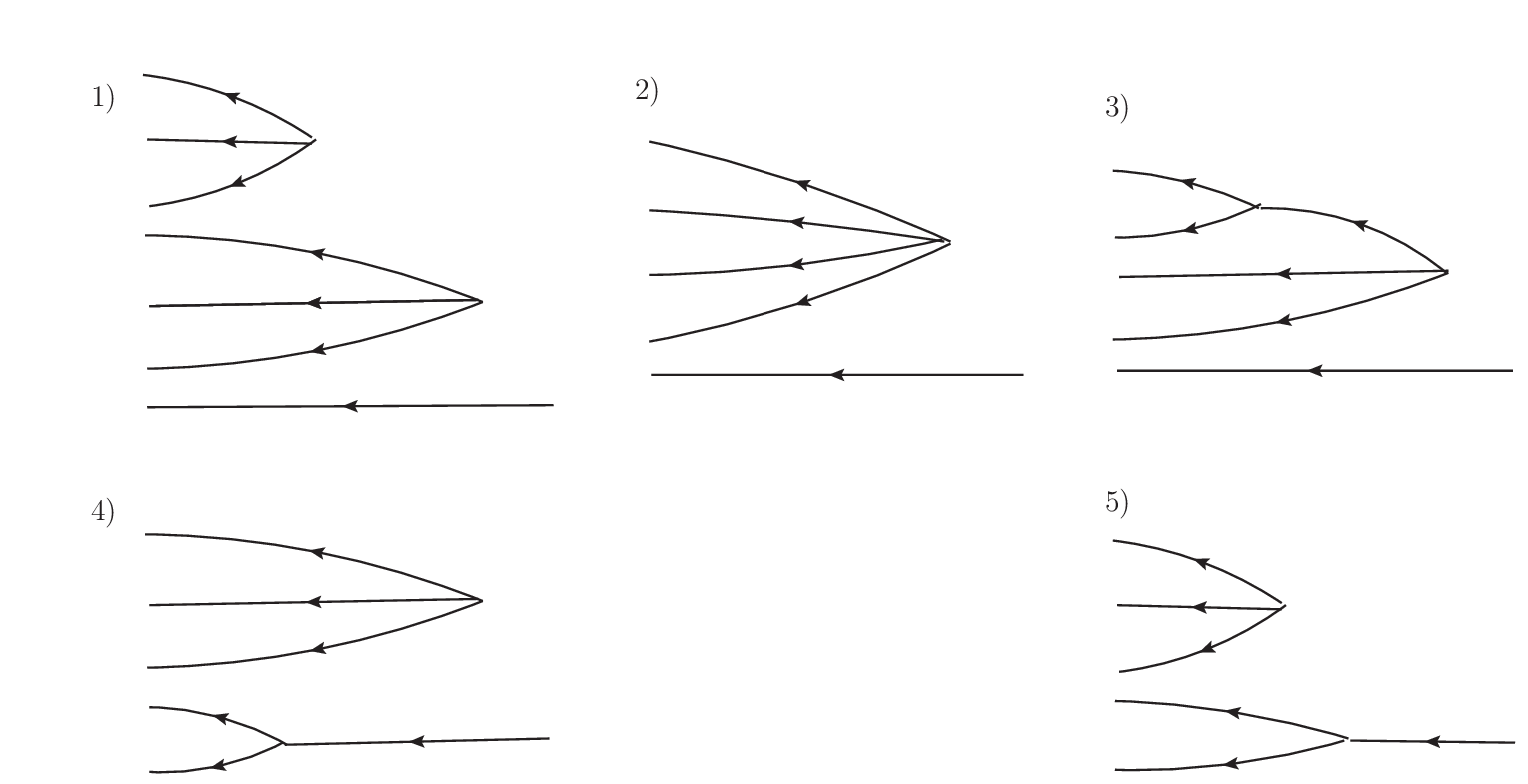}
\caption{The only contribution to $|\vp\rangle_2^7$ is shown in panel $(1)$.  The other panels show the contributions to $|\vp\rangle_2^5$.}\label{m257fig}
\end{figure}

The simplest Fock sector is the seven-meson sector, as the only contribution arises from
\beq
H_3^3|\vp\rangle_1^4=-\frac{m^2\lambda}{2}\pinv{3}{p_1}\cdots \frac{d^3\vp_4}{(2\pi)^3}\frac{|\vp,\vp_1,\vp_2,-\vp_1-\vp_2,\vp_3,\vp_4,-\vp_3-\vp_4\rangle_0}{\ovp 1+\ovp 2+\omega_{\vp_1+\vp_2}}
\eeq
and so
\beq
|\vp\rangle_2^7=\frac{m^2\lambda}{2}\pinv{3}{p_1}\cdots \frac{d^3\vp_4}{(2\pi)^3}\frac{|\vp,\vp_1,\vp_2,-\vp_1-\vp_2,\vp_3,\vp_4,-\vp_3-\vp_4\rangle_0}{\left(\ovp 1+\ovp 2+\omega_{\vp_1+\vp_2}\right)\left(\omega_{\vp_1+\vp_2}+\omega_{\vp_3+\vp_4}+\sum_{i=1}^4 \ovp i
\right)}.
\eeq
This contribution is shown in Fig.~\ref{m257fig}.

There are four contributions to $|\vp\rangle_2^5$
\bea
H_4^4|\vp\rangle_0^1&=&\frac{\lambda}{4}\pinv{3}{p_1}\pinv{3}{p_2}\pinv{3}{p_3}|\vp,\vp_1,\vp_2,\vp_3,-\vp_1-\vp_2-\vp_3\rangle_0\\
H_3^1|\vp\rangle_1^4&=&-\frac{9m^2\lambda}{4}\pinv{3}{p_1}\pinv{3}{p_2}\pinv{3}{p_3}\frac{|\vp,\vp_1,\vp_2,\vp_3,-\vp_1-\vp_2-\vp_3\rangle_0}{\omega_{\vp_1+\vp_2}\left(\ovp 1+\ovp 2 +\omega_{\vp_1+\vp_2}\right)}\nonumber\\
&&-\frac{3m^2\lambda}{4}\pinv{3}{p_1}\pinv{3}{p_2}\pinv{3}{p_3}\frac{|\vp_3,\vp-\vp_3,\vp_1,\vp_2,-\vp_1-\vp_2\rangle_0}{\ovp{}\left(\ovp 1+\ovp 2 +\omega_{\vp_1+\vp_2}\right)}\nonumber\\
H_3^3|\vp\rangle_1^2&=&-\frac{3m^2\lambda}{4}\pinv{3}{p_1}\pinv{3}{p_2}\pinv{3}{p_3}\frac{|-\vp_3,\vp+\vp_3,\vp_1,\vp_2,-\vp_1-\vp_2\rangle_0}{\ovp{}\left(\ovp 3 +\omega_{\vp+\vp_3}-\ovp{}\right)}.\nonumber
\eea

Altogether these lead to
\bea
|\vp\rangle_2^5&=&\frac{\lambda}{4}\pinv{3}{p_1}\cdots \frac{d^3\vp_3}{(2\pi)^3}\left[
\left(\frac{9m^2}{\omega_{\vp_1+\vp_2}\left(\ovp 1+\ovp 2 +\omega_{\vp_1+\vp_2}\right)}-1
\right)\frac{|\vp,\vp_1,\vp_2,\vp_3,-\vp_1-\vp_2-\vp_3\rangle_0}{\ovp 1+\ovp 2+\ovp 3+\omega_{\vp_1+\vp_2+\vp_3}}
\right.\nonumber\\
&&\hspace{-1.5cm}\left.+\frac{3m^2}{\ovp{}}\left(
\frac{1}{\left(\ovp 1+\ovp 2 +\omega_{\vp_1+\vp_2}\right)}
+\frac{1}{\left(\ovp 3 +\omega_{\vp+\vp_3}-\ovp{}\right)}
\right)\frac{|-\vp_3,\vp+\vp_3,\vp_1,\vp_2,-\vp_1-\vp_2\rangle_0}{\omega_{\vp+\vp_3}+\omega_{\vp_1+\vp_2}-\ovp{}
+\sum_{i=1}^3\ovp i}\right]\nonumber\\
&=&\frac{\lambda}{4}\pinv{3}{p_1}\cdots \frac{d^3\vp_3}{(2\pi)^3}\left[
\left(\frac{9m^2}{\omega_{\vp_1+\vp_2}\left(\ovp 1+\ovp 2 +\omega_{\vp_1+\vp_2}\right)}-1
\right)\frac{|\vp,\vp_1,\vp_2,\vp_3,-\vp_1-\vp_2-\vp_3\rangle_0}{\ovp 1+\ovp 2+\ovp 3+\omega_{\vp_1+\vp_2+\vp_3}}
\right.\nonumber\\
&&\hspace{-0.0cm}\left.+\frac{3m^2}{\ovp{}}\frac{|-\vp_3,\vp+\vp_3,\vp_1,\vp_2,-\vp_1-\vp_2\rangle_0}{\left(\ovp 1+\ovp 2 +\omega_{\vp_1+\vp_2}\right)\left(\ovp 3 +\omega_{\vp+\vp_3}-\ovp{}\right)}\right]
.
\eea

%\blu{
%I got
%\[
%\left\lvert \vec{p} \right\rangle_ {2}^{5} = \frac{g^{2}}{4}\int \frac{d^{3}p_ {1,2,3}}{(2\pi)^{9}} \,  \frac{\frac{9m^{2}}{\omega_ {2+3}(\omega_ {1}+\omega_ {2+3}+\omega_ {1+2+3})}-1}{\omega_ {1}+\omega_ {2}+\omega_ {3}+\omega_ {1+2+3}}\left\lvert \vec{p},\vec{p}_ {1,2,3,-1-2-3} \right\rangle_ {0} 
%+ \frac{3m^{2}\left\lvert \vec{p}_ {1,2,-1-2,-3},\vec{p}+\vec{p}_ {3} \right\rangle_ {0}}{\omega_ {p}(\omega_ {1}+\omega_ {2}+\omega_ {1+2})(\omega_ {3}+\omega_ {p+p_ {3}}-\omega_ {p})}
%\]
%but I guess they are equal to your answer. 
%}

There are five contributions to $|\vp\rangle_2^3$.  The first two arise from $H_4$
\beq
H_4^2|\vp\rangle_0^1=\frac{\lambda}{2\ovp{}}\pinv{3}{p_1}\pinv{3}{p_2}|\vp_1,\vp_2,\vp-\vp_1-\vp_2\rangle_0
-\frac{\delta m^2}{2}\pinv{3}{p_1}|\vp,\vp_1,-\vp_1\rangle_0.
\eeq
The others arise from $H_3$
\bea
H_3^{-1}|\vp\rangle_1^4&=&-\frac{9m^2\lambda}{4}\pinv{3}{p_1}\ppinv{3}{p}\frac{|\vp,\vp_1,-\vp_1\rangle_0}{\ovpp{}\omega_{\vp_1+\vpp}\left(\ovp 1+\ovpp{}+\omega_{\vp_1+\vpp}\right)}
\\
&&-\frac{9m^2\lambda}{4\ovp{}}\pinv{3}{p_1}\pinv{3}{p_2}\frac{|-\vp_1,-\vp_2,\vp+\vp_1+\vp_2\rangle_0}{\omega_{\vp_1+\vp_2}\left(\ovp 1+\ovp 2+\omega_{\vp_1+\vp_2} 
\right)}
\nonumber\\
H_3^{1}|\vp\rangle_1^2&=&-\frac{9m^2\lambda}{4\ovp{}}\pinv{3}{p_1}\pinv{3}{p_2}\frac{|-\vp_1,-\vp_2,\vp+\vp_1+\vp_2\rangle_0}{\omega_{\vp_1+\vp_2}\left(\omega_{\vp_1+\vp_2}+\omega_{\vp+\vp_1+\vp_2}-\ovp{} 
\right)}.\nonumber
\eea

\begin{figure}[htbp]
\centering
\includegraphics[width = 0.85\textwidth]{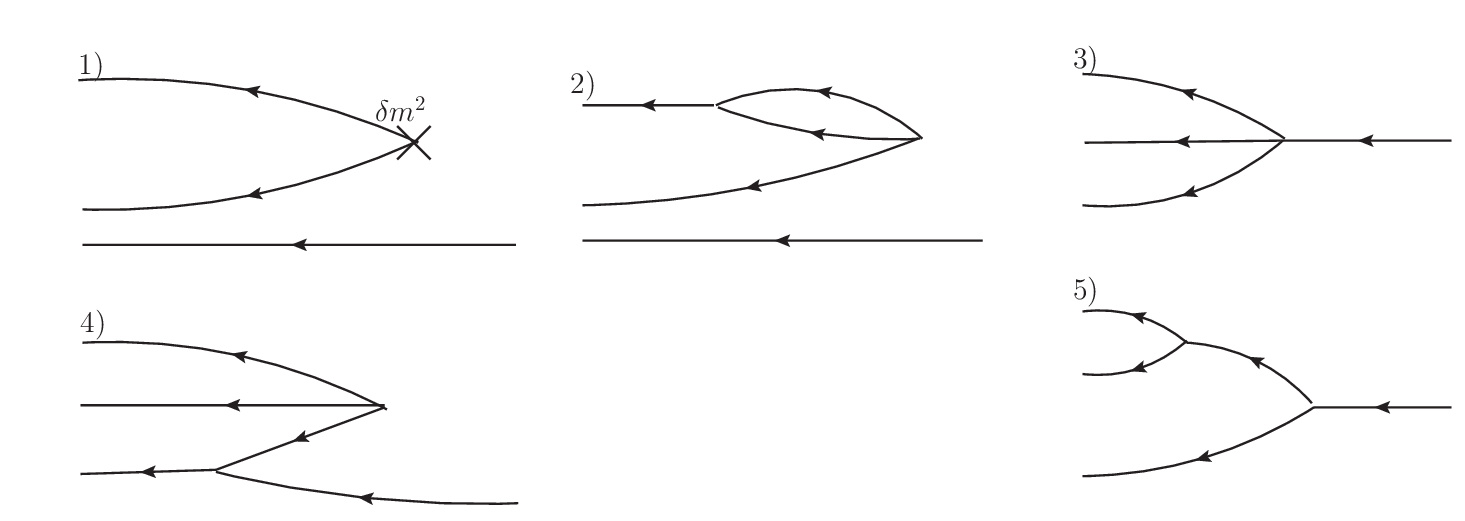}
\caption{These are the five contributions to $|\vp\rangle_2^3$, the part of the order $O(\lambda)$ correction to the Hamiltonian eigenstate $|\vp\rangle$ which lies in the three-meson Fock space.}\label{m23fig}
\end{figure}

Inverting $\ovp{}-H_2$ we find the leading three-meson contribution to the one-meson Hamiltonian eigenstate
\bea
|\vp\rangle_2^3&=&\pinv{3}{p_1}
\left(\frac{\delta m^2}{4} 
+\frac{9m^2\lambda}{8}\ppinv{3}{p}\frac{1}{\ovpp{}\omega_{\vp_1+\vpp}\left(\ovp 1+\ovpp{}+\omega_{\vp_1+\vpp}\right)}\right)
\frac{|\vp,\vp_1,-\vp_1\rangle_0}{\ovp{1}}\nonumber
\\
&&\hspace{-1cm}+\frac{\lambda}{2\ovp{}}\pinv{3}{p_1}\pinv{3}{p_2}\left[ \frac{9m^2}{2\omega_{\vp_1+\vp_2}}\left(\frac{1}{\ovp 1+\ovp 2+\omega_{\vp_1+\vp_2}}+\frac{1}{\omega_{\vp_1+\vp_2}+\omega_{\vp+\vp_1+\vp_2}-\ovp{}} 
\right)
-1
\right]\nonumber\\
&&\times\frac{|-\vp_1,-\vp_2,\vp+\vp_1+\vp_2\rangle_0}{\left(\ovp 1+\ovp 2+\omega_{\vp+\vp_1+\vp_2}-\ovp{}\right)}
%\nonumber\\
%&&\hspace{-1cm}+\frac{9m^2\lambda}{4\ovp{}}\pinv{3}{p_1}\pinv{3}{p_2}\frac{|-\vp_1,-\vp_2,\vp+\vp_1+\vp_2\rangle_0}{\omega_{\vp_1+\vp_2}\left(\ovp 1+\ovp 2+\omega_{\vp_1+\vp_2} 
%\right)\left(\ovp 1+\ovp 2+\omega_{\vp+\vp_1+\vp_2}-\ovp{}\right)}
%\nonumber\\
%&&+\frac{9m^2\lambda}{4\ovp{}}\pinv{3}{p_1}\pinv{3}{p_2}\frac{|-\vp_1,-\vp_2,\vp+\vp_1+\vp_2\rangle_0}{\omega_{\vp_1+\vp_2}\left(\omega_{\vp_1+\vp_2}+\omega_{\vp+\vp_1+\vp_2}-\ovp{} 
%\right)\left(\ovp 1+\ovp 2+\omega_{\vp+\vp_1+\vp_2}-\ovp{}\right)}\nonumber
\eea
which is shown in Fig.~\ref{m23fig}.  Note that the ultraviolet divergence on the first line is canceled if Eq.~(\ref{dma}) is satisfied.

\begin{figure}[htbp]
\centering
\includegraphics[width = 0.85\textwidth]{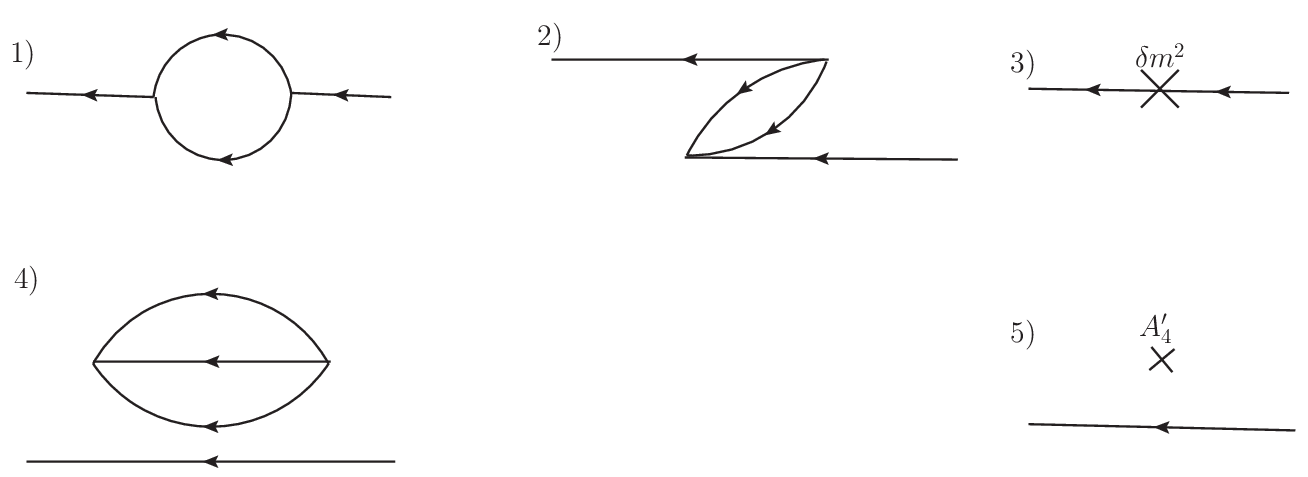}
\caption{These are the five contributions to $|\vp\rangle_2^1$.  The bottom two enjoy infrared divergences that cancel one another.  Our normalization condition implies that the total of all five terms vanishes, which allows us to calculate $\delta m^2$.}\label{m21fig}
\end{figure}

There are also five contributions to $|\vp\rangle_2^1$, shown in Fig.~\ref{m21fig}.  Two of these, shown in panels (4) and (5), suffer from infrared divergences, and so we will write them before $\dvx$ integration as
\bea
\ch_4^0(\vx)|\vp\rangle_0^1&\supset& A\p_4|\vp\rangle_0^1\label{dis}\\
\ch_3^{-3}|\vp\rangle_1^4&\supset&-\frac{3m^2\lambda}{8}\ppinv{3}{p_1}\ppinv{3}{p_2} \frac{1}{\ovpp 1\ovpp 2 \omega_{\vpp_1+\vpp_2}\left(\ovpp 1+\ovpp 2+\omega_{\vpp_1+\vpp_2}
\right)}
|\vp\rangle_0^1\nonumber
\eea
where the $\supset$ notation indicates that we have only considered the $c$-number term in $\ch_4$ and only certain contractions with $\ch_3$.  We will consider the other terms momentarily. 

The sum of the two terms in Eq.~(\ref{dis}) is exactly zero as a result of Eq.~(\ref{a4}).  Of course this is no surprise, the role of the counterterm $A_4$ is to cancel disconnected diagrams in the vacuum sector.  Note that the cancellation needs to be and is exact, as any mismatch would diverge when integrated over $\vx$.  In the soliton sector the cancellation will not be exact, even in the case of the domain wall ground state, but it will be $x_1$-dependent and so the integral over $x_1$ will converge.

The other three contributions, on the other hand, diverge in the ultraviolet
\bea
H_4^0|\vp\rangle_0^1&\supset&-\frac{\delta m^2}{2\ovp{}}|\vp\rangle_0
\\
H_3^{-3}|\vp\rangle_1^4&\supset&-\frac{9m^2\lambda}{8\ovp{}}\ppinv{3}{p}\frac{|\vp\rangle_0}{\ovpp{}\omega_{\vp+\vpp}\left(\ovp{}+\ovpp{}+\omega_{\vp+\vpp} \right)}
\nonumber\\
H_3^{-1}|\vp\rangle_1^2&=&-\frac{9m^2\lambda}{8\ovp{}}\ppinv{3}{p}\frac{|\vp\rangle_0}{\ovpp{}\omega_{\vp+\vpp}\left(\ovpp{}+\omega_{\vp+\vpp}-\ovp{} \right)}
\nonumber
\eea
where the $\supset$ symbol indicates that we are considering the terms corresponding to panels $(1-3)$ in Fig.~\ref{m21fig}, which we ignored above.  

These three terms also need to cancel precisely, as the right hand side of Eq.~(\ref{lamp}) vanishes in the one-meson sector because $(\ovp{}-H_2)|\vp\rangle_0=0$, or alternately because of our \hyperlink{terzo}{convention (III)} $|\vp\rangle_2^1=0$.  Adding the three contributions and setting the sum to zero we find
\beq
\delta m^2=-\frac{9m^2\lambda}{2}\ppinv{3}{p}\frac{\ovpp{}+\omega_{\vp+\vpp}}{\ovpp{}\omega_{\vp+\vpp}\left(\left(\ovpp{}+\omega_{\vp+\vpp}\right)^2-\ovp{}^2 \right)}=\left(-\frac{9}{8\pi^2}{\rm{ln}}\left(\frac{2\Lambda}{m}\right)+\frac{3\sqrt{3}}{16\pi}\right)m^2\lambda \label{dm}
\eeq
where $\vpp$ is integrated over a sphere of radius $\Lambda$.  This satisfies Eq.~(\ref{dma}) and so the coefficients are indeed finite.  Note also that it is independent of $\vp$, so that if the \hyperlink{second}{second} renormalization condition is applied at one value of $\vp_0$, it holds at all values.  This follows from the Lorentz covariance of the renormalization condition.
%\red{This depends on $\vp$.  The \hyperlink{second}{second} renormalization condition only holds at $\vp_0$, so we should impose $\vp=\vp_0$ here where $\vp_0$ can be chosen arbitrarily.  However, Lorentz invariance seems to say that if the second condition holds at one value of $\vp$, it should hold at all values.  But that would imply that the integral is independent of $\vp$.  Is it?  Anyway, even if it is in some sense, the cutoff will ruin the Lorentz invariance.  The cutoff chooses a frame, and $\delta m^2$ depends on that frame.  But the \hyperlink{second}{second} renormalization condition should still hold for all $\vp_0$ if it holds for one ... does it?}

Finally we turn to the \hyperlink{third}{third} renormalization condition.  The inner product is proportional to $|\vp_0\rangle_2^2$.  However, only odd meson numbers have appeared at this order.  Therefore, this renormalization condition is trivially satisfied.

\subsection{Order $O(\lambda^{3/2})$}

\subsubsection{Setup}

We have checked that all renormalization conditions can be satisfied at order $O(\lambda)$.  A similar calculation at order $O(\lambda^{3/2})$ would be somewhat lengthy and so we will perform it elsewhere.  For the purpose of testing the finiteness of the domain wall tension, we only need the asymptoic behavior of the divergent counterterm $\delta\sl$, which falls from the \hyperlink{third}{third} renormalization condition at order $O(\lambda^{3/2})$.  This renormalization condition concerns $|\vp\rangle_3^2$, to which we now turn.

At meson number two, the \hyperlink{second}{second} renormalization condition at order $O(\lambda^{3/2})$ is
\beq
H_5^1|\vp\rangle_0^1+H_4^{-2}|\vp\rangle_1^4+H_4^{0}|\vp\rangle_1^2+H_3^{-3}|\vp\rangle_2^5+H_3^{-1}|\vp\rangle_2^3+H_3^{1}|\vp\rangle_2^1=(\ovp{}-H_2)|\vp\rangle_3^2. \label{lam3}
\eeq

\subsubsection{Coupling Renormalization}

We are free to choose $\vp_0$, $\vp_1$ and $\vp_2$ in the \hyperlink{third}{third} renormalization condition.  Let us choose values such that neither $\vp_1$ nor $\vp_2$ is equal to $\vp_0$.  There will be contributions from components of $|\vp\rangle_3^2$ consisting of two mesons, one of which is $\vp$.  This choice means that, for the purpose of imposing the \hyperlink{third}{third} renormalization condition, we can drop all such contributions in Eq.~(\ref{lam3}).  In particular, we are not interested in the tadpole terms in $H_5$, which yield contributions of this form\footnote{Presumably the tadpole contributions would anyway cancel as a result of the \hyperlink{fourth}{fourth} renormalization condition, which fixes $\delta v_3$, but we will not impose this.}.  

Therefore we are only interested in the cubic term in $\ch_5(\vx)$
\beq
\ch_5(\vx)\supset m\sqrt{\frac{\lambda}{2}}\left(\frac{\delta m^2}{2m^2}+\frac{\delta\sl}{\sl} \right)\phi^3(\vx)
\eeq
which leads to
\beq
H_5^1\supset  \frac{3m}{4}\sqrt{\frac{\lambda}{2}}\left(\frac{\delta m^2}{m^2}+2\frac{\delta\sl}{\sl} \right)\pinv{3}{p_1}\pinv{3}{p_2} \frac{\Ad 1\Ad 2 A_{\vp_1+\vp_2}}{\omega_{\vp_1+\vp_2}}.
\eeq
One then finds the first contribution
\beq
H_5^1|\vp\rangle_0^1=\frac{3m}{4\ovp{}}\sqrt{\frac{\lambda}{2}}\left(\frac{\delta m^2}{m^2}+2\frac{\delta\sl}{\sl} \right)\pinv{3}{p_1}|-\vp_1,\vp+\vp_1\rangle_0^2. \label{dlcon}
\eeq
Our strategy to evaluate the ultraviolet divergent piece of $\delta\sl$ will be to ensure that this contribution cancels the ultraviolet divergent contributions from the other terms on the left hand side of Eq.~(\ref{lam3}).

\subsubsection{Reducible Ultraviolet Divergent Contributions}

First we turn to reducible divergences, involving insertions and loops on external legs of the three-point function.  These may be removed by amputating external legs.

In our application of the renormalization conditions at order $O(\lambda)$ we have already seen that two ultraviolet divergences cancel in the construction of $|\vp\rangle_2^1$.  After this cancellation, no more ultraviolet divergences remain in $H_3^1|\vp\rangle_2^1$ and so we will not consider this contribution further.

\begin{figure}[htbp]
\centering
\includegraphics[width = 0.85\textwidth]{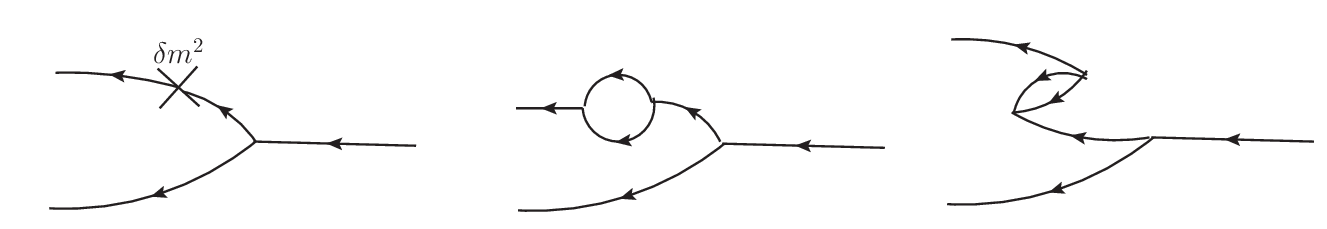}
\caption{These three contributions to $|\vp\rangle_3^2$ contain divergences that are reducible, in the sense that they can be amputated by cutting an external leg.  In these cases the external leg is outgoing.}\label{redfig}
\end{figure}

However, a related cancellation occurs between the following three contributions, drawn in Fig.~\ref{redfig}
\bea
H_4^{0}|\vp\rangle_1^2&\supset&-\frac{3m\delta m^2\sl}{2\sqrt{2}\ovp{}}\pinv{3}{p_1}\frac{|-\vp_1,\vp+\vp_1\rangle_0}{\ovp 1\left(\ovp 1+\omega_{\vp+\vp_1}-\ovp{}\right)}
\\
H_3^{-1}|\vp\rangle_2^3&\supset&-\frac{27m^3\lambda^{3/2}}{8\sqrt{2}\ovp{}}
\pinv{3}{p_1}\left[\ppinv{3}{p}\frac{1}{\omega_{\vp_1+\vpp}\ovpp{}\left(\omega_{\vp_1+\vpp}+\ovpp{}+\omega_{\vp+\vp_1}-\ovp{}\right)}\right]\nonumber\\
&&\times\frac{|-\vp_1,\vp+\vp_1\rangle_0}{\ovp 1\left(\omega_{\vp_1}+\omega_{\vp+\vp_1}-\ovp{}\right)} 
\nonumber\\
H_3^{-3}|\vp\rangle_2^5&\supset&-\frac{27m^3\lambda^{3/2}}{8\sqrt{2}\ovp{}}\pinv{3}{p_1}\left[\ppinv{3}{p}\frac{1}{\omega_{\vpp+\vp_1}\ovpp{}\left(\omega_{\vp+\vp_1}+2\omega_{\vp_1}+\ovpp{}+\omega_{\vp_1+\vpp}-\ovp{}\right)}\right]\nonumber\\
&&\times\frac{|-\vp_1,\vp+\vp_1\rangle_0}{\ovp 1\left(\omega_{\vp_1}+\omega_{\vp+\vp_1}-\ovp{}\right)} .
\nonumber
\eea
At high $\vpp=|\vpp|$, the quantities in square brackets diverge as $\ppinv{3}{p} 1/(2p^{\prime 3})$ and so, at each value of $\vp_1$, they cancel the divergence in the first term where $\delta m^2$ was given in Eq.~(\ref{dm}).  Note that the terms outside of the square brackets have the same $\vp$ and $\vp_1$ dependence, and so the cancellation applies at all values of $\vp$ and $\vp_1$.

\begin{figure}[htbp]
\centering
\includegraphics[width = 0.85\textwidth]{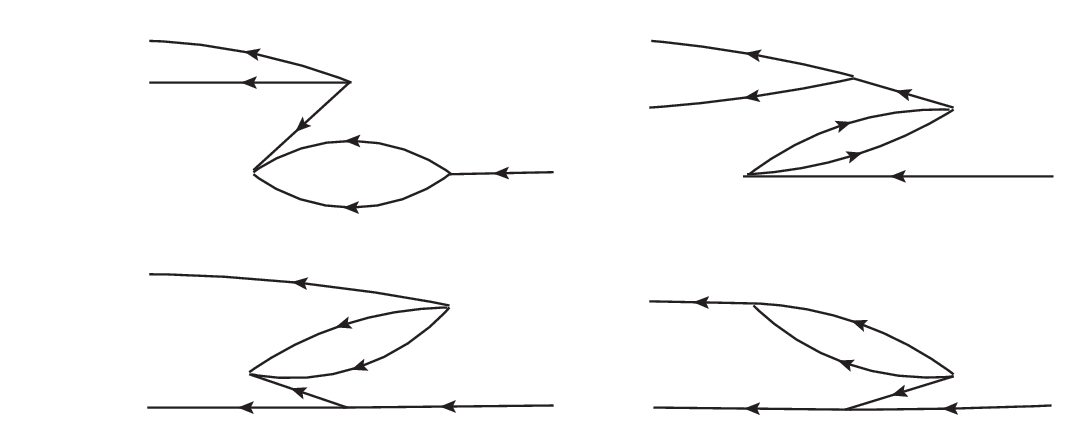}
\caption{These four contributions to $|\vp\rangle_3^2$ contain a loop, but are power counting finite because the $(E-H_2)^{-1}$ at a vertex which does not lie on the loop contributes a factor of $1/\vpp$.}\label{finfig}
\end{figure}

A similar triplet of divergences corresponds to reducible loops that arise at a lower order than the initial $\vp$ interaction.  These are all terms in $H_3^{-1}|\vp\rangle_2^3$.  Their sum is finite because, as we argued in the $O(\lambda)$ calculation, $|\vp\rangle_2^3$ is finite as a result of Eq.~(\ref{dma}).  There are also four related diagrams with a reducible loop, shown in Fig.~\ref{finfig}.  They contain four powers of $p\p$ in the denominator and so are finite.

\begin{figure}[htbp]
\centering
\includegraphics[width = 0.85\textwidth]{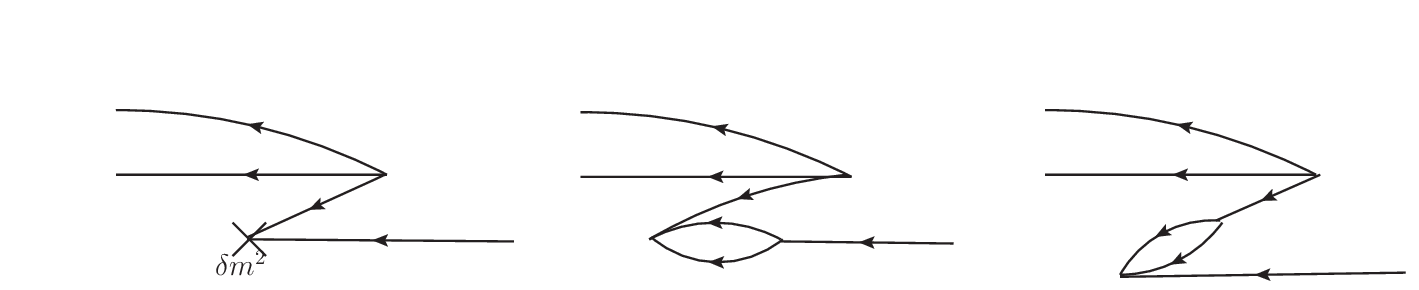}
\caption{As in Fig.~\ref{redfig}, but these divergences lie on the incoming leg.}\label{red2fig}
\end{figure}

Finally, three terms shown in Fig.~\ref{red2fig} contain reducible loops or counterterm insertions on the momentum $\vp$ leg
\bea
H_4^{-2}|\vp\rangle_1^4&\supset&-\frac{3m\delta m^2\sl}{4\sqrt{2}\ovp{}^2}\pinv{3}{p_1}\frac{|-\vp_1,\vp+\vp_1\rangle_0}{\left(\omega_{\vp_1}+\omega_{\vp+\vp_1}+\ovp{}\right)} .
\nonumber\\
H_3^{-3}|\vp\rangle_2^5&\supset&-\frac{27m^3\lambda^{3/2}}{16\sqrt{2}\ovp{}^2}\pinv{3}{p_1}\left[\ppinv{3}{p}\frac{1}{\ovpp{}\omega_{\vp+\vpp}}\left(\frac{1}{\ovpp{}+\omega_{\vp+\vpp}+\ovp 1+\omega_{\vp+\vp_1}} \right.\right.\nonumber\\
&&\left.\left.
+\frac{1}{\ovp 1+\omega_{\vp+\vp_1}+\ovpp{}+\omega_{\vp+\vpp}}
\right)
\right]%\\&&\times
\frac{|-\vp_1,\vp+\vp_1\rangle_0}{\left(\ovp{}+\ovp{1}+\omega_{\vp+\vp_1}\right)}.
\nonumber
\eea
The last two terms are equal.  Adding them, the term in the square bracket diverges as $\ppinv{3}{p} 1/(2p^{\prime 3})$.  Therefore again, at each value of $\vp_1$, they cancel the divergence in the first term as a result of Eq.~(\ref{dm}).  As expected, reducible loops do not contribute to the divergence in $\delta\sl$.

\subsubsection{Irreducible Ultraviolet Divergent Contributions}

\begin{figure}[htbp]
\centering
\includegraphics[width = 0.65\textwidth]{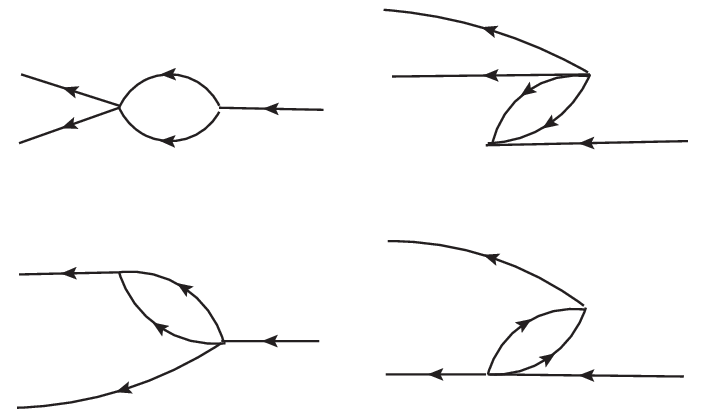}
\caption{These are the divergent contributions of interest to $|\vp\rangle_3^2$.  The third renormalization condition implies that they must cancel the counterterm in Eq.~(\ref{dlcon}), which corresponds to a diagram of a single, degree three vertex.  This allows us to fix $\delta\sl$.}\label{irredfig}
\end{figure}

There are four contributions with irreducible loops shown in Fig.~\ref{irredfig}
\bea
H_4^0|\vp\rangle_1^2&\supset&\frac{9m\lambda^{3/2}}{8\sqrt{2}\ovp{}}\pinv{3}{p_1}\left[ \ppinv{3}{p}\frac{1}{\ovpp{}\omega_{\vpp+\vp}\left(\ovpp{}+\omega_{\vp+\vpp}-\ovp{} 
\right)}\right]|-\vp_1,\vp+\vp_1\rangle_0
\\
H_3^{-3}|\vp\rangle_2^5&\supset&\frac{9m\lambda^{3/2}}{8\sqrt{2}\ovp{}}\pinv{3}{p_1}\left[ \ppinv{3}{p}\frac{1}{\ovpp{}\omega_{\vpp+\vp}\left(\ovp 1+\ovpp{}+\omega_{\vp+\vp_1}+\omega_{\vp+\vpp} 
\right)}\right]|-\vp_1,\vp+\vp_1\rangle_0
\nonumber\\
H_3^{-1}|\vp\rangle_2^3&\supset&\frac{9m\lambda^{3/2}}{4\sqrt{2}\ovp{}}\pinv{3}{p_1}\left[ \ppinv{3}{p}\frac{1}{\ovpp{}\omega_{\vpp+\vp_1}\left(\ovpp{}+\omega_{\vp_1+\vpp}+\omega_{\vp+\vp_1}-\ovp{} 
\right)}\right]|-\vp_1,\vp+\vp_1\rangle_0
\nonumber\\
H_4^{-2}|\vp\rangle_1^4&\supset&\frac{9m\lambda^{3/2}}{4\sqrt{2}\ovp{}}\pinv{3}{p_1}\left[ \ppinv{3}{p}\frac{1}{\ovpp{}\omega_{\vpp+\vp_1}\left(\ovp 1+\ovpp{}+\omega_{\vp_1+\vpp} 
\right)}\right]|-\vp_1,\vp+\vp_1\rangle_0.
\nonumber
\eea
The terms in square brackets each diverge in the ultraviolet as $\ppinv{3}{p}1/(2p^{\prime 3})$ and so, up to finite terms, these four contributions sum to
\beq
\frac{27m\lambda^{3/2}}{8\sqrt{2}\ovp{}}\pinv{3}{p_1} \left[ \ppinv{3}{p}\frac{1}{p^{\p 3}}\right]|-\vp_1,\vp+\vp_1\rangle_0=
\frac{27m\lambda^{3/2}}{16\sqrt{2}\pi^2\ovp{}}{\rm{ln}}(\Lambda)\pinv{3}{p_1}|-\vp_1,\vp+\vp_1\rangle_0.
\eeq

This must cancel the contribution (\ref{dlcon}) from the counterterms.  Therefore, up to finite corrections
\beq
\left(\frac{\delta m^2}{m^2}+2\frac{\delta\sl}{\sl}\right)=-\frac{9\lambda}{2}\ppinv{3}{p}\frac{1}{p^{\p 3}}=-\frac{9\lambda}{4\pi^2}{\rm{ln}}(\Lambda).
\eeq
Using the value of $\delta m^2$ reported in Eq.~(\ref{dm}) we find, up to a finite term, the coupling constant renormalization
\beq
\delta\sl=-\frac{9\lambda^{3/2}}{16\pi^2}{\rm{ln}}(\Lambda).
\eeq
Inserting this and Eq.~(\ref{dm}) into Eq.~(\ref{v1}), we see that $\delta v_1$ is finite.  This is a necessary condition for the operator $\dv$ to exist.

\section{The Domain Wall Sector} \label{wallsez}

\subsection{Definitions}

Consider a solution
\beq
\phi(\vx,t)=F(\vx)=f(x_1)
\eeq
to the classical field equations with the renormalized parameters, corresponding to the first line of Eq.~(\ref{hhp}).  In other words
\beq
\partial_1^2 f(x_1)=-\frac{m^2}{2}f(x_1)+\lambda f^3(x_1). \label{eom}
\eeq
We will be interested in the domain wall solution
\beq
f(x_1)=\frac{m}{\sqrt{2\lambda}}\tanh{\left(\frac{mx_1}{2}\right)}.
\eeq
We may also consider a vacuum solution $f=\pm m/\sqrt{2\lambda}$ but in that case the following reasoning would lead us to the results of the previous sections.

Now construct the displacement operator
\beq
\df=\exp{-i\dvx F(\vx) \pi(\vx)}
\eeq
which satisfies
\beq
\phi(\vx)\df=\df\left(\phi(\vx)+F(\vx)\right)\hsp [\pi(\vx),\df]=0.
\eeq
Acting as in the vacuum sector, we may use the displacement operator to construct a domain wall sector Hamiltonian
\beq
H\p[\phi,\pi]=\df^\dag \hat H[\phi,\pi]\df=\hat H[\phi+F,\pi]\hsp H\p=\dvx :\ch\p(\vx):_m.
\eeq

The result looks rather like Eq.~(\ref{chsb})
\bea
\ch(\vx)&=&\frac{\pi^2(\vx)+\left(\partial_1(\phi(\vx)+f(x_1))\right)^2+\sum_{i=2}^3\left(\partial_i \phi(\vx)\right)^2}{2}\label{hp}\\
&&+\frac{\lambda (\phi(\vx)+f(x_1))^4-m^2 (\phi(\vx)+f(x_1))^2}{4}+A
\nonumber\\
&&\hspace{-2cm}+\frac{\left(-2\sl\delta\sl+(\delta\sl)^2 \right) (\phi(\vx)+f(x_1))^4+\left(\delta m^2+6\left(\sl-\delta\sl\right)^2 I\right)  (\phi(\vx)+f(x_1))^2}{4}.\nonumber
\eea

Why have we constructed a new Hamiltonian?  Because if any state $|\psi\rangle$ is in the vacuum sector, then $\df\dv^\dag|\psi\rangle$ will be in the domain wall sector.  If
\beq
H\df\dv^\dag|\psi\rangle=E\df\dv^\dag|\psi\rangle \label{nonp}
\eeq
then
\beq
H\p|\psi\rangle=E|\psi\rangle. \label{pert}
\eeq
However the Fock space of nonperturbative states $\df\dv^\dag|\psi\rangle$ consisting of the domain wall plus a finite number of mesons is in one to one correspondence with the ordinary perturbative Fock space of states $|\psi\rangle$.  Thus by using $H\p$ instead of $H$ we can find the domain wall states and their energies using ordinary perturbation theory.  

Said differently, the domain wall states $\df\dv^\dag|\psi\rangle$ are intrinsically nonperturbative, because $\df$ and $\dv$ contain an exponentiated $1/\sl$, and they satisfy the difficult equation (\ref{nonp}).  However, to find the spectrum of $H$ in the domain wall sector, it is sufficient to solve the equation (\ref{pert}) for the perturbative states $|\psi\rangle$.  In other words, the domain wall sector consists of perturbative eigenstates of $H\p$.

\subsection{Decomposing the Domain Wall Hamiltonian}

Since we have already calculated the counterterms, now to calculate the domain wall tension to order $O(\lambda^0)$ we need only consider the terms in the Hamiltonian up to order $O(\lambda^0)$.  Let us now decompose $H\p$
\beq
H\p=\sum_{i=0}^\infty H\p_i\hsp H\p_i=\dvx :\ch\p_i(\vx):_m
\eeq
where $H\p_i$ is of order $O(\lambda^{i/2-1})$.

At leading order (\ref{hp}) reduces to
\beq
\ch\p_0(\vx)=\frac{f^{\prime 2}(x_1)}{2}+\frac{\lambda f^4(x_1)-m^2f^2(x_1)}{4}+\frac{m^4}{16\lambda}
\eeq
where we have used the value of $A_0$ found in Eq.~(\ref{a0}).  Of course this is just the energy density of the $\phi^4$ kink in a theory with the bare parameters replaced by the renormalized parameters, and so it integrates to the classical kink mass
\beq
\rho=\sum_{i=0}^\infty\rho_i\hsp \rho_i=\int dx_1 :\ch\p_{2i}(\vx):_m\hsp
\rho_0=\frac{m^3}{3\lambda}.
\eeq
In 3+1 dimensions, $\rho$ is not the kink mass but rather the tension of the domain wall, which extends in the $x_2$ and $x_3$ directions.  We conclude that at leading order the tension $\rho_0$ of the domain wall agrees with the classical result, which is not a surprise.

At order $O(1/\sl)$, after integrating by parts, Eq.~(\ref{hp}) yields
\beq
\ch\p_1(\vx)= \phi(\vx)\left[-\partial_1^2 f(x_1)
+\lambda f^3(x_1)-\frac{m^2}{2} f(x_1)
\right]=0
\eeq
where in the last step we used the truncated classical equation of motion (\ref{eom}).  Of course this is not entirely trivial, as the full classical equation of motion contains the bare parameters.  However, the difference will not arise until order $O(\sl)$ and so will not affect the one-loop tension.  We may therefore expect an unpleasant surprise in the calculation of the two-loop tension, but this will have to await further work.

Finally we turn to the domain wall Hamiltonian density at order $O(\lambda^0)$
\bea
\ch\p_2(\vx)&=&\frac{\pi^2(\vx)+\sum_{i=1}^3\left(\partial_i \phi(\vx)\right)^2+(3\lambda f^2(x_1)-m^2/2)\phi^2(\vx)}{2}\\
&&+\frac{f^2(x_1)\delta m^2-2f^4(x_1)\sl\delta\sl
}{4}+\frac{m^4}{8\lambda}\left(\frac{\delta\sl}{\sl}-\frac{\delta m^2}{m^2}\right)\nonumber
\eea
where we have used the value of $A_2$ reported in Eq.~(\ref{a2}).

The first line is an operator, let us call it $\ch^{\prime A}_2(\vx)$ while the second is a $c$-number that we will call $\ch^{\prime B}_2(\vx)$.  We will find that both contain ultraviolet divergences. 

\subsection{Divergences from the $c$-number Term}

To obtain the next correction $\rho_1$ to the tension, we need to integrate both terms over $x_1$
\beq
\rho_1=\rho_1^A+\rho_1^B.
\eeq
Let us begin with the $c$-number term
\bea
\rho_1^B&=&\frac{\delta m^2}{m^2}\int dx_1 \left(\frac{m^2 f^2(x_1)}{4}-\frac{m^4}{8\lambda}\right)+\frac{\delta\sl}{\sl}\int dx_1\left(-\frac{\lambda f^4(x_1)}{2}+ \frac{m^4}{8\lambda}\right)\nonumber\\
&=&-\frac{m^3}{2\lambda}\frac{\delta m^2}{m^2}+\frac{2m^3}{3\lambda}\frac{\delta\sl}{\sl}=\frac{3m^3}{16\pi^2}{\rm{ln}}(\Lambda)+{\rm{finite}}.\label{b}
\eea
We will drop the finite term in the last step because we have not calculated the finite contributions to $\delta\sl$.

We believe that in Ref.~\cite{erice}, when Coleman stated that the energy density of a coherent state corresponding to a soliton is infinite beyond 1+1 dimensions, he was referring to this divergence.  Indeed, the tension of the coherent domain wall state $\df\dv^\dag\ovac_0$ is $\rho_0+\rho^B_1$, which we have just shown is divergent.  In the next section we will construct what we believe is the correct domain wall state, which is a squeezed coherent state where the squeeze exactly cancels this divergence.

\subsection{Constructing the Domain Wall State}

We are interested in the lowest energy eigenstate of the Hamiltonian $H\p_2$, and the corresponding eigenvalue of the tension operator
\beq
\rho_1^A=\frac{1}{2}\int dx_1 \left( :\pi^2(\vx):_m+\sum_{i=1}^3:\left(\partial_i \phi(\vx)\right)^2:_m+\V2:\phi^2(\vx):_m
\right)
\eeq
where have introduced the shorthand
\beq
\V2=3\lambda f^2(x_1)-m^2/2.
\eeq
%\gre{all the notation $\V2$ Better to be $ V^{2}[\sqrt{\lambda}f(\vx)]$ to keep consistent with ref[23]?}
This problem was solved in Ref.~\cite{me2d}.  Let us review the solution here.

The classical equation of motion corresponding to $H\p_2$ is
\beq
\left(-\partial_t^2+\partial_i^2\right)\phi(\vx,t)=\V2 \phi(\vx,t).
\eeq
A basis of constant, negative frequency solutions is
\beq
\phi(\vx,t)=\g_{k_1k_2k_3}(\vx)e^{-i\omega_{k_1k_2k_3}t}\hsp
\g_{k_1k_2k_3}(\vx)=\g_{k_1}(x_1)e^{-i(k_2x_2+k_3x_3)}. \label{phis}
\eeq
Here the functions $\g_{k}(x)$ are solutions to the P\"oschl-Teller potential
\bea
\g_k(x)&=&\frac{e^{-ikx}}{\ok{} \sqrt{m^2+4k^2}}\left[2k^2-m^2+(3/2)m^2\sech^2(m x/2)-3im k\tanh(m x/2)\right]\nonumber\\
\g_S(x)&=&\frac{\sqrt{3 m}}{2}\tanh(m x/2)\sech(m x/2)\hsp
\g_B(x)=-\sqrt{\frac{{3m}}{8}}\sech^2(m x/2).\label{nmode}
\eea
Note that in (\ref{phis}) the abstract index $k_1$ runs over the real numbers and also the two discrete values $B$ and $S$ whereas in the first line of (\ref{nmode}), the same letter $k$ only runs over the real numbers.
The indices $B$ and $S$ represent the zero mode and shape mode of the (1+1)-dimensional $\phi^4$ kink.  The frequencies $\omega$ are defined to be
\beq
\omega_{B k_2 k_3}=\sqrt{k_2^2+k_3^2}\hsp \omega_{S k_2 k_3}=\sqrt{\frac{3m^2}{4}+k_2^2+k_3^2}\hsp \omega_{k_1k_2 k_3}=\sqrt{m^2+k_1^2+k_2^2+k_3^2} \label{omk}
\eeq
where $k_1$, like $k_2$ and $k_3$, runs over the real numbers.

The equation of motion for $H\p_2$ implies that the functions $\g_k(x)$ satisfy the Sturm-Liouville equations
\beq
\V2\g_k(x)=\left(\partial_x^2+\omega^2_{k 0 0}\right)\g_k(x)
\eeq
or equivalently
\beq
\V2\g_{k_1 k_2 k_3}(\vx)=\left(\nabla^2+\omega^2_{k_1 k_2 k_3}\right)\g_{k_1k_2k_3}(\vx). \label{sl}
\eeq

Following Ref.~\cite{cahill76} we decompose the field $\phi(\vx)$ and its momentum $\pi(\vx)$ in normal modes
\bea
\phi(\vx)&=&\ppink{3}\g_{\vk}(\vx)\phi_\vk\nonumber\\
\pi(\vx)&=&\ppink{3}\g_{\vk}(\vx)\pi_\vk
\eea
%\bea
%\phi(\vx)&=&\phi_0\g_{B00}(x_1)+\ppink{3}\g_{\vk}(\vx)\left(B^{\ddag}_\vk+\frac{B_{-\vk}}{2\omega_\vk}\right)\nonumber\\
%\pi(\vx)&=&\pi_0\g_{B00}(x_1)+i\ppink{3}\g_{\vx}(\vx)\left(\omega_{\vk} B^{\ddag}_\vk-\frac{B_{-\vk}}{\omega_\vk}\right)\hsp
%B^{\ddag}_\vk=\frac{B^{\dag}_\vk}{2\omega_\vk}
%\eea
where $\dint$ sums over real triplets $(k_1,k_2,k_3)$ and also $(B,k_1,k_2)$ and $(S,k_2,k_3)$ where $B$ and $S$ are the abstract indices representing the zero mode and shape mode.  %However it does not sum over the case of the zero frequency $(B,0,0)$.

Integrating by parts, $\ch^{\prime A}_2(\vx)$ can be simplified using the Sturm-Liouville equations (\ref{sl})
\bea
\ch^{\prime A}_2(\vx)&=&\frac{\pi^2(\vx)+\phi(\vx)\left[\V2-\nabla^2\right]\phi(\vx)}{2} \label{chpa}\\
&=&\frac{1}{2}\ppink{3}\ppinkp{3}\g_{\vk}(\vx)\g_{\vkp}(\vx)\left[\pi_{\vk}\pi_{\vkp}+\omega^2_{\vk_2}\phi_{\vk}\phi_{\vkp}\right].\nonumber
\eea
The Sturm-Liouville completeness relation can be used to fix the normalizations
\beq
\int dx \g_{k}(x)\g_{k\p}(x)=2\pi \delta(k+k\p)\hsp
\int dx \g_{S}(x)\g_{S}(x)=\int dx \g_B(x)\g_B(x)=1
\eeq
where other combinations yield zero by the orthogonality of the solutions.  Inserting this into (\ref{chpa}) we find
\bea
\int dx_1 \ch^{\prime A}_2(\vx)&=&\frac{1}{2}\ppin{k_1}\int\frac{dk_2dk_3dk\p_2dk\p_3}{(2\pi)^4}
e^{-i(k_2+k\p_2)x_2-i(k_3+k\p_3)x_3}\label{dena}\\
&&\times\left[\pi_{k_1k_2k_3}\pi_{-k_1k\p_2k\p_3}+\omega^2_{k_1k\p_2k\p_3}\phi_{k_1k_2k_3}\phi_{-k_1k\p_2k\p_3}\right].\nonumber
\eea

What is the eigenvector of $H^{\prime A}_2$ with minimum eigenvalue?  Integrating (\ref{dena}) over $x_2$ and $x_3$, we see that $k_2=-k\p_2$ and $k_3=-k\p_3$ and so this reduces to a sum of quantum harmonic oscillators.  The ground state $\vac_0$ is therefore the state annihilated by $\pi_\vk- i\omega_{\vk}\phi_\vk$.  The normal ordering of an operator quadratic in the fields only shifts the operator by a $c$-number, and so does not affect the eigenvectors or the ordering of their eigenvalues, and so does not alter this conclusion.

%The orthogonality of the plane waves then implies
%\bea
%\dvx \g_{\vk_1}(\vx)\g_{\vk_2}(\vx)&=&(2\pi)^3 \delta^3(\vk_1+\vk_2)\\
%\dvx \g_{Sk_{11}k_{12}}(\vx)\g_{Sk_{21}k_{22}}(\vx)&=&\dvx \g_{Sk_{11}k_{12}}(\vx)\g_{Sk_{21}k_{22}}(\vx)=(2\pi)^2\delta(\vk_{11}+\vk{21})\delta(\vk_{12}+\vk_{22}).\nonumber
%\eea

\subsection{The Tension In General}

The $A$ contribution to the tension is the eigenvalue of
\beq
\rho^{A}_1=\int dx_1 :\ch_2^{\prime A}(\vx):_m
\eeq
acting on $\vac_0$.  We will now turn to the evaluation of this contribution.

Let us define the operators
\beq
B_{\vk}=\frac{\phi_\vk}{2}+i\frac{\pi_\vk}{2\omega_{\vk}}
\eeq
that annihilate $\vac_0$ and as usual we will define the combination
\beq
B^\ddag_{\vk}=\frac{B^\dag_{\vk}}{2\omega_{\vk}}.
\eeq
Note that these are not defined in the case $\vk=(B00)$ and so in that case we will instead adopt the notation
\beq
\phi_{\vk=(B00)}=\phi_0\hsp\pi_{\vk=(B00)}=\pi_0
\eeq
and state that $\pi_0\vac_0=0$.  We will see that the following expressions are infrared finite and so the contributions from the small $k$ limit vanish, allowing us to safely ignore these operators.  This is in contrast with the 1+1 dimensional kink, in which the majority of the kink mass arises from the zero mode.

If the Hamiltonian $H\p_2$ were normal ordered with respect to the $B$ operators, so that each $B^\ddag$ appears on the left, then $H\p_2$ would annihilate $\vac_0$.  In that case the tension would be given by $\rho^B$ which diverges, and in fact is equal to that of a coherent state.  However it is normal ordered with respect to the $A$ operators, and it is this difference which will remove the divergence.

As normal ordering bilinears in fields yields a $c$-number, the difference between these normal orderings contributes a constant to the energy, which in fact is our tension $\rho_1^A$.

Combining the decompositions of the fields in normal modes and plane waves, one finds
\bea
\phi_{k_1k_2k_3}&=&\dvx \g_{-k_1}(x_1) e^{i(k_2x_2+k_3x_3)}\phi(\vx)\\
&=&\dvx \g_{-k_1}(x_1) e^{i(k_2x_2+k_3x_3)}\pinv{3}{p} e^{-i\vp\cdot\vx}\left(A^\ddag_{\vp}+\frac{A_{-\vp}}{2\ovp{}}\right)
\nonumber\\
&=&\pin{p_1}\tilde{\g}_{-k_1}(p_1)\left(A^\ddag_{p_1k_2k_3}+\frac{A_{-p_1-k_2-k_3}}{2\omega_{p_1k_2k_3}}\right)\nonumber\\
\pi_{k_1k_2k_3}&=&i\pin{p_1}\tilde{\g}_{-k_1}(p_1)\left(\omega_{p_1k_2k_3}A^\ddag_{p_1k_2k_3}-\frac{A_{-p_1-k_2-k_3}}{2}\right)\nonumber
\eea
where $\tilde{\g}$ is the Fourier transform of $\g$.

This implies
\bea
\omega^2_{k_1k\p_2k\p_3}:\phi_{k_1k_2k_3}\phi_{-k_1k\p_2k\p_3}:_m&=&\pin{p_1}\pin{p\p_1}\tilde{\g}_{-k_1}(p_1)\tilde{\g}_{k_1}(p\p_1)\omega^2_{k_1k\p_2k\p_3}\nonumber\\
&&\hspace{-1cm}\times\left(A^\ddag_{p_1k_2k_3}A^\ddag_{p\p_1k\p_2k\p_3}+A^\ddag_{p_1k_2k_3}\frac{A_{-p\p_1-k\p_2-k\p_3}}{2\omega_{p\p_1k\p_2k\p_3}}\right.\nonumber\\
&&\hspace{-1cm}\left.+A^\ddag_{p\p_1k\p_2k\p_3}\frac{A_{-p_1-k_2-k_3}}{2\omega_{p_1k_2k_3}}+\frac{A_{-p_1-k_2-k_3}}{2\omega_{p_1k_2k_3}}\frac{A_{-p\p_1-k\p_2-k\p_3}}{2\omega_{p\p_1k\p_2k\p_3}}\right)
\nonumber\\
:\pi_{k_1k_2k_3}\pi_{-k_1k\p_2k\p_3}:_m&=&
\pin{p_1}\pin{p\p_1}\tilde{\g}_{-k_1}(p_1)\tilde{\g}_{k_1}(p\p_1)\omega_{p_1k_2k_3}\omega_{p\p_1k\p_2k\p_3}\nonumber\\
&&\hspace{-1cm}\times\left(-A^\ddag_{p_1k_2k_3}A^\ddag_{p\p_1k\p_2k\p_3}+A^\ddag_{p_1k_2k_3}\frac{A_{-p\p_1-k\p_2-k\p_3}}{2\omega_{p\p_1k\p_2k\p_3}}\right.\nonumber\\
&&\hspace{-1cm}\left.+A^\ddag_{p\p_1k\p_2k\p_3}\frac{A_{-p_1-k_2-k_3}}{2\omega_{p_1k_2k_3}}-\frac{A_{-p_1-k_2-k_3}}{2\omega_{p_1k_2k_3}}\frac{A_{-p\p_1-k\p_2-k\p_3}}{2\omega_{p\p_1k\p_2k\p_3}}\right). \label{aa}
\eea

Now the normal ordering symbol has disappeared, because we have enforced the normal ordering.  All that remains is to act these operators on $\vac_0$.  We know that this is annihilated by the $B$ operators, and so we need to use a Bogoliubov transformation to change the $A$ operators into $B$ operators
\bea
A^\ddag_{p_1k_2k_3}&=&\frac{1}{2}\dvx e^{i(p_1x_1+k_2x_2+k_3x_3)}\left(\phi(\vx)-i\frac{\pi(\vx)}{\omega_{p_1k_2k_3}}\right)\\
&=&\frac{1}{2}\dvx e^{i(p_1x_1+k_2x_2+k_3x_3)}\ppinkp{3}\g_{k\p_1}(x_1)e^{-i(k\p_2x_2+k\p_3x_3)}\nonumber\\
&&\times\left(B^\ddag_{k\p_1k\p_2k\p_3}+\frac{B_{-k\p_1k\p_2k\p_3}}{2\omega_{k\p_1k\p_2k\p_3}} 
+\frac{\omega_{k\p_1k\p_2k\p_3}}{\omega_{p_1k_2k_3}}B^\ddag_{k\p_1k\p_2k\p_3}-\frac{B_{-k\p_1k\p_2k\p_3}}{2\omega_{p_1k_2k_3}}
\right)\nonumber\\
%&=&\frac{1}{2}\ppinkp{3}\tilde\g_{k\p_1}(x_1)(2\pi)^2\delta(k_2-k\p_2)\delta(k_3-k\p_3)\nonumber\\
%&&\times\left[\left(1+\frac{\omega_{k\p_1k\p_2k\p_3}}{\omega_{p_1k_2k_3}}\right)B^\ddag_{k\p_1k\p_2k\p_3}+\left(1-\frac{\omega_{k\p_1k\p_2k\p_3}}{\omega_{p_1k_2k_3}}\right)\frac{B_{-k\p_1k\p_2k\p_3}}{2\omega_{k\p_1k\p_2k\p_3}} \right]\nonumber\\
&=&\frac{1}{2}\ppin{k\p_1}\tilde\g_{k\p_1}(-p_1)\left[\left(1+\frac{\omega_{k\p_1k_2k_3}}{\omega_{p_1k_2k_3}}\right)B^\ddag_{k\p_1k_2k_3}+\left(1-\frac{\omega_{k\p_1k_2k_3}}{\omega_{p_1k_2k_3}}\right)\frac{B_{-k\p_1k_2k_3}}{2\omega_{k\p_1k_2k_3}} 
\right]\nonumber\\
\frac{A_{-p_1-k_2-k_3}}{2\omega_{p_1k_2k_3}}&=&\frac{1}{2}\ppin{k\p_1}\tilde\g_{k\p_1}(-p_1)\left[\left(1-\frac{\omega_{k\p_1k_2k_3}}{\omega_{p_1k_2k_3}}\right)B^\ddag_{k\p_1k_2k_3}+\left(1+\frac{\omega_{k\p_1k_2k_3}}{\omega_{p_1k_2k_3}}\right)\frac{B_{-k\p_1k_2k_3}}{2\omega_{k\p_1k_2k_3}} 
\right].\nonumber
\eea

Consider a term such as $AA$ in Eq.~(\ref{aa}).  The contribution to the tension comes from the difference between the plane wave normal ordering $::_m$ and the normal mode normal ordering $::_b$, which puts all $B$ to the right of $B^\ddag$.  Therefore it comes from the $B$ term in the first $A$ and the $B^\ddag$ term in the second $A$.  Acting on $\vac_0$, it is therefore equal to the commutator of the two
\bea
A^\ddag_{p_1k_2k_3}A^\ddag_{p\p_1k\p_2k\p_3}-:A^\ddag_{p_1k_2k_3}A^\ddag_{p\p_1k\p_2k\p_3}:_b&=&
\frac{1}{4}\ppin{k_1}\tilde\g_{k_1}(-p_1)\ppin{k\p_1}\tilde\g_{k\p_1}(-p\p_1)\\
&&\hspace{-5cm}\ \ \times\left(1-\frac{\omega_{k_1k_2k_3}}{\omega_{p_1k_2k_3}}\right) \left(1+\frac{\omega_{k_1k\p_2k\p_3}}{\omega_{p\p_1k\p_2k\p_3}}\right) 
\left[\frac{B_{-k_1 k_2k_3}}{2\omega_{k_1k_2k_3}},B^\ddag_{k\p_1k\p_2k\p_3} 
\right]\nonumber\\
&&\hspace{-5cm}=\frac{1}{8}\ppin{k_1}\frac{\tilde \g_{k_1}(-p_1)\tilde \g_{-k_1}(-p\p_1)}{\omega_{\vk}}\left(1-\frac{\omega_{k_1k_2k_3}}{\omega_{p_1k_2k_3}}\right) \left(1+\frac{\omega_{k_1k_2k_3}}{\omega_{p\p_1k_2k_3}}\right) \nonumber\\
&&\hspace{-5cm}\ \ \times(2\pi)^2\delta(k\p_2-k_2)\delta(k\p_3-k_3)\nonumber
\eea
and similarly
\bea
A^\ddag_{p_1k_2k_3}\frac{A_{-p\p_1-k\p_2-k\p_3}}{2\omega_{p\p_1k\p_2k\p_3}}-:A^\ddag_{p_1k_2k_3}\frac{A_{-p\p_1-k\p_2-k\p_3}}{2\omega_{p\p_1k\p_2k\p_3}}:_b\,&=&\,\,\frac{1}{8}\ppin{k_1}\frac{\tilde \g_{k_1}(-p_1)\tilde \g_{-k_1}(-p\p_1)}{\omega_{\vk}}\\
&&\hspace{-5cm}\times\left(1-\frac{\omega_{k_1k_2k_3}}{\omega_{p_1k_2k_3}}\right) \left(1-\frac{\omega_{k_1k_2k_3}}{\omega_{p\p_1k_2k_3}}\right)(2\pi)^2\delta(k\p_2-k_2)\delta(k\p_3-k_3)\nonumber\\
A^{\ddag}_{p\p_1k\p_2k\p_3}\frac{A_{-p_1-k_2-k_3}}{2\omega_{p_1k_2k_3}}-:A^\ddag_{p\p_1k\p_2k\p_3}\frac{A_{-p_1-k_2-k_3}}{2\omega_{p_1k_2k_3}}:_b&=&\frac{1}{8}\ppin{k_1}\frac{\tilde \g_{k_1}(-p_1)\tilde \g_{-k_1}(-p\p_1)}{\omega_{\vk}}\nonumber\\
&&\hspace{-5cm}\times\left(1-\frac{\omega_{k_1k_2k_3}}{\omega_{p_1k_2k_3}}\right) \left(1-\frac{\omega_{k_1k_2k_3}}{\omega_{p\p_1k_2k_3}}\right)(2\pi)^2\delta(k\p_2-k_2)\delta(k\p_3-k_3)\nonumber\\
\frac{A_{-p_1-k_2-k_3}}{2\omega_{p_1k_2k_3}}\frac{A_{-p\p_1-k\p_2-k\p_3}}{2\omega_{p\p_1k\p_2k\p_3}}-:\frac{A_{-p_1-k_2-k_3}}{2\omega_{p_1k_2k_3}}\frac{A_{-p\p_1-k\p_2-k\p_3}}{2\omega_{p\p_1k\p_2k\p_3}}:_b&=&\frac{1}{8}\ppin{k_1}\frac{\tilde \g_{k_1}(-p_1)\tilde \g_{-k_1}(-p\p_1)}{\omega_{\vk}}\nonumber\\
&&\hspace{-5cm}\times\left(1+\frac{\omega_{k_1k_2k_3}}{\omega_{p_1k_2k_3}}\right) \left(1-\frac{\omega_{k_1k_2k_3}}{\omega_{p\p_1k_2k_3}}\right)(2\pi)^2\delta(k\p_2-k_2)\delta(k\p_3-k_3).\nonumber
\eea
Inserting these into Eq.~(\ref{aa}) we find that the contributions to the tension are
\bea
\omega^2_{k_1k\p_2k\p_3}\left(:\phi_{k_1k_2k_3}\phi_{-k_1k\p_2k\p_3}:_m-:\phi_{k_1k_2k_3}\phi_{-k_1k\p_2k\p_3}:_b\right)
&=&\frac{1}{4}\pin{p_1}\pin{p\p_1}\tilde{\g}_{-k_1}(p_1)\tilde{\g}_{k_1}(p\p_1)\nonumber\\
&&\hspace{-6cm}\ \ \times \omega^2_{\vk}
\ppin{k\p_1}{\tilde \g_{k\p_1}(-p_1)\tilde \g_{-k\p_1}(-p\p_1)}\left(\frac{2}{\omega_{k\p_1k_2k_3}}-\frac{1}{\omega_{p_1k_2k_3}}-\frac{1}{\omega_{p\p_1k_2k_3}}\right)
\nonumber\\
&&\hspace{-6cm}\ \ \times(2\pi)^2\delta(k\p_2-k_2)\delta(k\p_3-k_3)
\nonumber\\
:\pi_{k_1k_2k_3}\pi_{-k_1k\p_2k\p_3}:_m-:\pi_{k_1k_2k_3}\pi_{-k_1k\p_2k\p_3}:_b&=&\frac{1}{4}\pin{p_1}\pin{p\p_1}\tilde{\g}_{-k_1}(p_1)\tilde{\g}_{k_1}(p\p_1)\nonumber\\&&\hspace{-6cm}\ \ \times 
\ppin{k\p_1}{\tilde \g_{k\p_1}(-p_1)\tilde \g_{-k\p_1}(-p\p_1)}\left(2\omega_{k\p_1k_2k_3}-{\omega_{p_1k_2k_3}-\omega_{p\p_1k_2k_3}}{}\right)\nonumber\\
&&\hspace{-6cm}\ \ \times {}(2\pi)^2\delta(k\p_2-k_2)\delta(k\p_3-k_3).
\nonumber
\eea

This in turn must be inserted into the tension (\ref{dena})
\bea
\int dx_1 \left(:\ch^{\prime A}_2(\vx):_m-:\ch^{\prime A}_2(\vx):_b\right)&=&\frac{1}{8}\ppink{3}\pin{p_1}\pin{p\p_1}\tilde{\g}_{-k_1}(p_1)\tilde{\g}_{k_1}(p\p_1)\\
&&\hspace{-6cm}\times
\ppin{k\p_1}{\tilde \g_{k\p_1}(-p_1)\tilde \g_{-k\p_1}(-p\p_1)}\left[2\omega_{k\p_1k_2k_3}-\omega_{p_1k_2k_3}-\omega_{p\p_1k_2k_3}+\frac{2\omega_\vk^2}{\omega_{k\p_1k_2k_3}}-\frac{\omega^2_\vk}{\omega_{p_1k_2k_3}}-\frac{\omega^2_\vk}{\omega_{p\p_1k_2k_3}} 
\right]\nonumber
\eea
%\left[:\pi_{\vk}\pi_{-\vk}:_m-:\pi_{\vk}\pi_{-\vk}:_b\right.\\
%&&\left.+\omega^2_{\vk}:\phi_{\vk}\phi_{-\vk}:_m-:\phi_\vk\phi_{-\vk}:_b\right].\nonumber
%\eea
where we remind the reader that the $::_b$ term annihilates $\vac_0$ as it consists of terms of the form $B^\ddag B$.

Notice that no summand in the square bracket has both $p_1$ and $p_1\p$.  Thus, whichever is missing may be integrated, using the completeness relation for the $\tilde\g$ to yield a $2\pi\delta(k_1+k\p_1)$, or a Kronecker $\delta$ when $k$ is a discrete mode.  This can be used to perform the $k\p_1$ integration, and so we find
\bea
\int dx_1 :\ch^{\prime A}_2(\vx):_m\vac_0&=&
\frac{1}{4}\ppink{3}\pin{p_1}|\tilde{\g}_{-k_1}(p_1)|^2
%\\&&\hspace{-0cm}\times
\left[2\omega_{k\p_1k_2k_3}-\omega_{p_1k_2k_3}-\frac{\omega^2_\vk}{\omega_{p_1k_2k_3}} 
\right]\vac_0\nonumber\\
&=&-\frac{1}{4}\ppink{3}\pin{p_1}|\tilde{\g}_{-k_1}(p_1)|^2\frac{\left(\omega_\vk-\omega_{p_1k_2k_3}\right)^2}{\omega_{p_1k_2k_3}}\vac_0.
\eea

We conclude that the corresponding contribution to the tension is
\beq
\rho_1^A=-\frac{1}{4}\ppink{3}\pin{p_1}|\tilde{\g}_{-k_1}(p_1)|^2\frac{\left(\omega_\vk-\omega_{p_1k_2k_3}\right)^2}{\omega_{p_1k_2k_3}}.
\eeq
This is a higher-dimensional generalization of the formula for a kink mass presented in Ref.~\cite{cahill76}.

\subsection{The Tension in this Model}

The logarithmic ultraviolet divergence in $\rho_1^A$ arises from the continuum modes, for which $k_1$ may be arbitrarily large.  Thus we will restrict our attention to those modes.  In Ref.~\cite{mekink} the Fourier transform
\beq
\tilde{\g}_k(p)=\frac{2k^2-m^2}{\ok{}\sqrt{m^2+4k^2}}2\pi\delta(p+k)+\frac{6\pi p}{\ok{}\sqrt{m^2+4k^2}} \csch\left(\frac{\pi (p+k)}{m}\right)
\eeq
was computed. The Dirac $\delta$ term does not contribute to the tension, because each occurrence is multiplied by a zero.  

The continuum contribution is therefore
\bea
\rho_1^A&=&-\pin{k_1}\pin{p_1}\frac{9\pi^2 p_1^2}{(m^2+k_1^2)(m^2+4k_1^2)}\csch^2\left(\frac{\pi (p_1-k_1)}{m}\right)\int\frac{dk_2dk_3}{(2\pi)^2}\frac{\left(\omega_\vk-\omega_{p_1k_2k_3}\right)^2}{\omega_{p_1k_2k_3}}\nonumber\\
&=&-\frac{3\pi}{2}\pin{k_1}\pin{p_1}\frac{p_1^2\sqrt{m^2+k_1^2}}{(m^2+4k_1^2)}\csch^2\left(\frac{\pi (p_1-k_1)}{m}\right)\nonumber\\
&&\times\left(2-3\sqrt{\frac{m^2+p_1^2}{m^2+k_1^2}}+\left(\frac{m^2+p_1^2}{m^2+k_1^2}\right)^{3/2}\right).
\eea
%\gre{there may be $\csch{\frac{\pi(p_1-k_1}{m}}$ inside (4.42),because $g_{-k}(p)$ have sign flip}
As a result of the $\csch$, at large $k_1$ or $p_1$, the support of the tension is at 
\beq
\epsilon=p_1-k_1
\eeq
%\gre{there may be $ p_1-k_1 $ according to above note, however the final result is even about the$ k_1$,so the sign flip will not change the finial result}
of order $O(m)$.  In particular, this is much smaller than $p_1$ or $k_1$.  We thus expand to second order in $\epsilon/k_1$ to obtain the limiting behavior at large $k_1$
\beq
2-3\sqrt{\frac{m^2+p_1^2}{m^2+k_1^2}}+\left(\frac{m^2+p_1^2}{m^2+k_1^2}\right)^{3/2}\sim\frac{3\epsilon^2k_1^2}{(m^2+k_1^2)^2}\sim\frac{3\epsilon^2}{k_1^2}
\eeq
and so
\bea
\rho^A_1&\sim& -\frac{9\pi}{8}\pin{k_1}\frac{1}{|k_1|}\pin{\epsilon}\epsilon^2\csch^2\left(\frac{\pi\epsilon}{m}\right)\nonumber\\
&=&-\frac{3m^3}{16\pi}\pin{k_1}\frac{1}{|k_1|}=-\frac{3m^3}{16\pi^2}{\rm{ln}}(\Lambda) \label{a}
\eea
where we have included a factor of two from the fact that $|k_1|$ is large but the sign must be summed over.  This divergent negative energy cancels, up to finite contributions, the positive energy divergence from $\rho_1^B$ reported in Eq.~(\ref{b}).  This cancellation is our main result.

\section{Remarks}

49 years ago, Coleman noted that solitons above 1+1 dimensions cannot be described by coherent states, because their energy densities would be infinite.  We have reproduced this ultraviolet divergence in Eq.~(\ref{b}).  

However, solitons are not described by coherent states even in 1+1 dimensions.  At order $e^{-1/\sl}$ they are described by coherent states, but there is a correction of order unity.  This correction changes the true perturbative vacuum $\ovac_0$, which is annihilated by the $A_\vp$ operators, to the squeezed state $\vac_0$ which is annihilated by the $B_\vk$ operators.  Physically this is because the normal modes are turned off in the ground state, instead of the plane waves.  These states are related by a Bogoliubov transformation, and so $\vac_0$ is a squeezed state.  The soliton ground state is not the coherent state $\df\ovac_0$, rather it is the squeezed coherent state $\df\vac_0$, plus perturbative corrections that are suppressed by powers of $\sl$.

This squeeze contributes a divergent negative energy density, calculated in Eq.~(\ref{a}), which exactly cancels the divergence resulting from the displacement operator in the coherent state, leaving a finite energy density and even a finite tension.

Thus, at one loop, we feel that we have succeeded in constructing the quantum state corresponding to a domain wall in the 3+1 dimensional double-well theory. 

At this point the generalization to multiple loops is still not obvious.  In principle, it may depend on our choices of operator $\df$, although it may well be that any change in $\df$ could be absorbed into a change in $\vac$.  Also, our somewhat unconventional Schrodinger picture renormalization conditions may lead to divergences.

Our derivation is not the same as that of Ref.~\cite{noi4dlett}.  One crucial difference is that, in that reference, the operator $\df$ was constructed using the bare mass and coupling.  This made the construction very fast, but it is not clear that such an operator exists.  In the present note the operator $\df$ was constructed using the renormalized mass and coupling.  This operator exists and leads to the same results.  In fact, the two operators only differ by terms which are suppressed by powers of $\sl$, and so changing the construction of $\df$ is compensated by the different perturbative corrections that arise.

Domain walls have a number of applications, featuring in many cosmological and phenomenological models \cite{okun}, and displaying a rich dynamics \cite{bp21,bp23}.  In 2+1 dimensions, the one-loop quantum correction to the domain wall tension has been calculated in Ref.~\cite{zar98}.  Keeping the finite terms in the calculation above, one would arrive at the domain wall tension in 3+1 dimensions.  The two-loop quantum correction in 2+1 dimensions has a similar divergence structure to the one-loop correction in 3+1 dimensions, and so the same approach may be used to calculate this two-loop correction.  To move to two loops in 3+1 dimensions, one needs to understand the role of quadratic divergences and field renormalization.  

The Skyrme model \cite{skyrmion} is nonrenormalizable, but at one loop is a sensible theory.  It describes nuclei at large $N$ \cite{wittskyrme}, but is often thought to be a poor description of nuclei at $N=3$ because the predicted binding energies are an order of magnitude too large.  There has recently been a surge of interest in one-loop quantum corrections to Skyrmions \cite{qs1,qs2,qs3,qs4,qs5,qs6}.  Recent calculations \cite{bjarkequant} of some contributions to the one loop energy suggest that these quantum corrections largely solve the binding energy problem, making the model phenomenologically viable.  Thus, it would be of interest to apply our formalism to the Skyrme model, where a full calculation of one-loop corrections is in principle possible, albeit numerically challenging beyond the smallest nuclei.

Our motivation for studying ultraviolet divergences is not only to allow the methods of Refs.~\cite{mekink} and \cite{me2loop} to be applied to more than one spatial dimension, but also to allow a more general field content.  For example, recently there has been a renaissance in studies of kinks in interesting models with fermions \cite{ferm1,ferm2,ferm3,ferm4,ferm5,ferm6} to which we will turn before our ultimate goal of constructing the 't Hooft-Polyakov monopole state.  Needless to say, we are also interested in solitons in theories with yet more challenging field content \cite{zhong1,zhong2,zhong3}.

\section* {Acknowledgement}

\noindent
JE is supported by NSFC MianShang grants 11875296 and 11675223.  HL is supported by the Ministry of Education, Science, Culture and Sport of the Republic of Armenia under the Postdoc-Armenia Program, grant number 24PostDoc/2‐1C009. JE and HL are supported by the Ministry of Education, Science, Culture and Sport of the Republic of Armenia under the Remote Laboratory Program, grant number 24RL-1C047. BYZ is supported by the Young Scientists Fund of the National Natural Science Foundation of China (Grant No. 12305079).

\end{document}